\documentclass[12pt]{article}
\usepackage{amsfonts,amssymb,amsmath,mathrsfs,latexsym,bbm,dsfont,amsthm}
\usepackage{color}
\usepackage{graphicx}
\usepackage[scriptsize,nooneline,hang]{caption}
\usepackage[hang,nooneline,scriptsize]{subfigure}
\usepackage{rotating}
\begin{document}

\begin{center}
{\large \bf  Regular Black Holes as Particle Accelerators}
\end{center}

\vskip 5mm

\begin{center}
{\Large{Parthapratim Pradhan\footnote{E-mail: pppradhan77@gmail.com}}}
\end{center}

\vskip  0.5 cm

{\centerline{\it Department of Physics}}
{\centerline{\it Vivekananda Satabarshiki Mahavidyalaya}}
{\centerline{\it Manikpara, West Midnapur}}
{\centerline{\it West Bengal~721513, India}}

\vskip 1cm

\begin{abstract}
We investigate the possibility of arbitrarily high energy in the center
of mass(CM) frame of colliding particles in the vicinity of the infinite red-shift
surface of the spherically symmetric, static charged  regular
black holes (Bardeen black hole, Ay\'{o}n-Beato and Garc\'{i}a black hole,
and Hayward black hole). We show that the CM energy of colliding
particles near the infinite red-shift surface of the \emph{extreme} regular black
holes are arbitrarily large while the non-extreme regular black holes have the
finite energy. We also compute the equation of innermost stable circular
orbit(ISCO), marginally bound circular orbit(MBCO) and circular photon
orbit(CPO) of the above regular black holes, which are most relevant to
black hole accretion disk theory.
\end{abstract}

\section{Introduction}
Recently, Ba\~{n}ados, Silk and West(hereafter BSW) \cite{bsw}  demonstrated that
particles falling freely from rest exterior of a rotating extremal black hole can
produce an infinite amount of high CM energy. In a semi-realistic
setup, this energy could be higher than the Planckian energy, so that one might think
extremal black hole could be act as super high energy particle accelerator.  Soon
the appearance of this work in the literature several practical objections to the
proposed mechanism has been raised.
Particularly in \cite{berti}, the authors have shown that there is an astrophysical
bound i.e. maximal spin, back reaction effect and gravitational radiation etc. on
that CM energy due to the Thorn's bound \cite{thorn} i.e. $a=0.998M$
($M$ is the mass and $a=\frac{J}{M}$ is the angular momentum per unit mass of the
black hole or Kerr parameter). Also Jacobson et al. \cite{jacob} showed that CM energy
in the near extremal situation for  Kerr black hole is $\frac{E_{cm}}{2m_{0}}\sim \frac{2.41}{(1-a)^{1/4}}$.
Lake \cite{lake} found that the CM energy at the inner horizon of a static Reissner-Nordstr{\o}m(RN) black hole and Kerr black holes are limited. Grib et al. \cite{grib} investigated the CM energy using the multiple scattering mechanism.
Also in \cite{grib1}, the same authors computed the CM energy of particle collisions
in the ergo-sphere of the Kerr black hole. The collision in the ISCO particles was investigated by the Harada et al. \cite{harada} for Kerr black hole.  Liu et al. \cite{liu} studied the BSW effect for Kerr-Taub-NUT(Newman, Unti, Tamburino) space-time and proved that the CM energy depends upon  both the Kerr parameter ($a$) and the NUT  parameter ($n$) of the space-time. Li et al. \cite{li} proved that the Kerr-AdS  black hole space-time could be act as a particle accelerator. Studies were done by Zhong et al.\cite{zong} for RN-de-Sitter  black hole and found that infinite energy in the CM frame near the cosmological horizon.  Said et al. \cite{said} studied the particle
accelerations and collisions in the back ground of a cylindrical black holes.
In \cite{piran}, the authors suggested that using collisional Penrose process the emitted massive  particles can only be gain $\sim 30$ percentage of the initial rest energy of
the in-falling particles.

In \cite{wei1}, the authors discussed the CM energy for the Kerr-Newman black hole.
For Kerr-Sen black hole, the CM energy is diverging  also discussed in \cite{wei}.
It was discussed in \cite{zaslav} regarding the BSW process for spherically symmetric
RN black hole.  Zhu et al. \cite{zhu} showed that general stationary charged
black holes as charge particle accelerators.  In \cite{frolov}, the author demonstrated that a weakly magnetized black hole may behave as a particle accelerators. The BSW process was considered for regular black hole, BTZ black hole and Einstein-Maxwell-Dilation-Axion  in \cite{huss}. The effect of particle accelerations and collisions on the near horizon surface of some black holes was discussed in \cite{sharif}.

Also McWilliams \cite{mc} showed that the black holes are neither particle accelerators nor dark matter probes.  Galajinsky \cite{gala} also showed that the CM  energy in the context of the near horizon geometry of the extremal Kerr black holes and proved that the CM energy is finite for any value of the particle parameters. Tursunov et al. \cite{tur} studied the particle accelerations and collisions in case of black string. Fernando \cite{fernando} has studied the possibility of high CM energy of two particles colliding near the infinite red-shift surface of a charged black hole in string theory. Studies of BSW effect were performed in \cite{patil5,patil7} for the naked singularity case of different space-times.

In a recent work \cite{pp3}, we  have demonstrated that an extremal static, spherically symmetric string black hole could act as a particle accelerator with arbitrarily high energy when two uncharged particles falling freely from rest to infinity on the near horizon. It was also shown there that, the CM energy of collision is independent of the extreme  fine tuning of the angular momentum parameter of the colliding particles.

In a present work, our objective is to show an analogous effect of particle collisions with a high CM energy is also possible when the black hole is precisely \emph{extremal regular} static and spherically symmetric black hole.

Traditional black holes like Schwarzschild black hole, RN black hole, Kerr black hole
and Kerr-Newman black holes have possessed a curvature singularity. Whereas the regular black hole does not possess any curvature singularity.  This is the main difference between the traditional black hole and the regular black hole. On the other hand, it is also important to understand the final state of gravitational collapse of initially regular configurations we need to study the global regularity of black hole solutions.  Broadly speaking, for traditional black holes the curvature invariants
$R, R_{ab}R^{ab}, R_{abcd}R^{abcd}$ blows up at $r=0$ in the spacetime manifold, while for regular black holes the curvature invariants do not blow up everywhere in the spacetime manifold including at $r=0$. In this sense it is said ``regular black hole'' or ``non-singular black hole''.

We have considered here a number of regular black holes like: the Bardeen black
hole \cite{bard}, Ay\'{o}n-Beato \& Garc\'{i}a (ABG) black hole \cite{abg1} and Hayward black hole \cite{hay}.  The special features of these black holes are they have satisfied the weak energy condition(WEC) and the energy-momentum tensor should have the symmetry  $T_{00}=T_{11}$. Our goal here is to see what happens the BSW effect for these black holes in the \emph{extremal limit}.

Circular geodesic motion in the equatorial plane is of fundamental interest in
black hole physics as well as in accretion disk physics to determine the important
features of the spacetime. Circular orbits with $r>r_{ISCO}$ are stable, while those
with $r<r_{ISCO}$ are not. Keplerian circular orbits exist in the region $r>r_{ph}$, with $r_{ph}$ being the circular photon orbit. Bound circular orbits exist in the region $r>r_{mb}$, with $r_{mb}$ being the marginally bound circular orbit, and stable circular orbits exist for $r>r_{ISCO}$, with $r_{ISCO}$. We have calculated these orbits for the above regular black holes.  These orbits
are very crucial in black hole physics as well as in astrophysics because they determine
important information on the back ground geometry.

There are a number of references we would like to mention here which discuss the some
interesting properties of the regular black holes. Firstly, Ansoldi \cite{ansol} gives a good review about the regular black hole. Balart \cite{lb} studied the Brown-York quasi-local energy and  Komar energy at the horizon which is called Bose-Dadhich identity \cite{bose}, and proved that this identity does not satisfy for regular black holes.
In \cite{nora}, the author showed that Smarr's formula do not satisfy in
case of the regular Bardeen and ABG space-time. In \cite{bala}, the authors
have studied the weak energy condition of the regular black hole.
Recently, Garc\'{i}a et al. \cite{eva} have discussed the complete geodesic structure of ABG spacetime by using Weierstrass elliptic functions. Finally, Eiroa et al. \cite{eiro} have discussed the gravitational lensing effect of the regular Bardeen black hole.

The paper is organized as follows. In section 2, we compute the CM energy of the
particle collision near the infinite red-shift surface of the Bardeen Space-time.
In section 3, we describe the BSW effect of the ABG space-time. In section 4, we
calculate the $E_{cm}$ for the Hayward black hole and finally we summarize the
results in section 5.

\section{CM energy of the collision near the horizon of the Bardeen Space-time:}
In this section, we  shall first examine the CM energy of collision for two neutral
particles falling freely from rest at infinity in the horizon of a Bardeen black hole-
the \emph{first} regular (singularity-free) black hole model in general
relativity(GR). The space-time has an interesting feature: it is
interpreted as the gravitational field of a nonlinear monopole, i.e., as
a magnetic solution to Einstein field equations coupled to a non-linear electrodynamics.
To compute the CM energy near the horizon of the regular Bardeen black hole, we first need to know the geodesic structure of this black hole and four velocity of the particles.

\subsection{Equatorial circular orbit in the Bardeen space-time:}

Thus the line element for the Bardeen space-time\cite{bard,abg,borde,ansol,zhou} is
given by
\begin{eqnarray}
ds^2=-{\cal F}(r)dt^{2}+\frac{dr^{2}}{{\cal F}(r)}
+r^{2}\left(d\theta^{2}+\sin^{2}\theta d\phi^{2}\right) ~.\label{sph}
\end{eqnarray}
where the function ${\cal F}(r)$ is defined by
\begin{eqnarray}
{\cal F}(r) &=& 1- \frac{2mr^2}{(r^2+g^2)^{\frac{3}{2}}}.
\end{eqnarray}
where $m$ is the mass of the black hole and $g$ is the monopole charge
of the non-linear self gravitating magnetic field.
We can see the behaviour of the function ${\cal F}(r)$ for different
values of $g$ in the following plot.

\begin{figure}[t]
\begin{center}
{\includegraphics[width=0.45\textwidth]{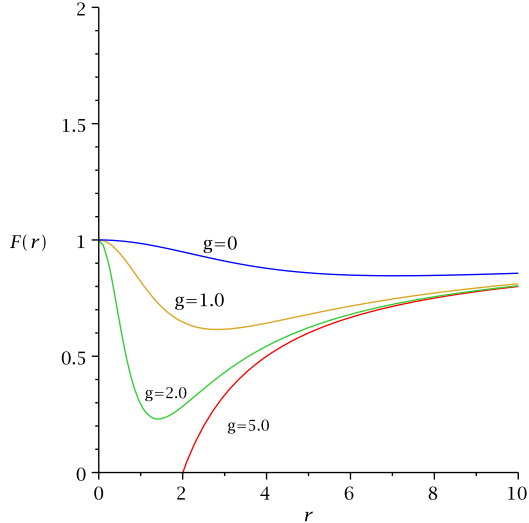}}
\end{center}
\caption{The figure shows the variation  of ${\cal F}(r)$  with $r$ for
different values of $g$.  Here, $m=1$, $g=0$ is the Schwarzschild case.
\label{bdf1}}
\end{figure}


It may be noted that this metric function behaves asymptotically i.e
at $r\rightarrow \infty$ as \cite{borde}
\begin{eqnarray}
{\cal F}(r) \sim 1-\frac{2m}{r}+\frac{3mg^2}{r^3}+O(\frac{1}{r^{5}}) ~.\label{asymbd}
\end{eqnarray}

Furthermore, if $r\rightarrow 0$, the metric function behaves as the de-Sitter
space-time, that is
\begin{eqnarray}
{\cal F}(r) \sim 1-\frac{2m}{g^3}r^2~.\label{asymds}
\end{eqnarray}

The Bardeen black hole has an event horizon $(r_{+})$ which occur at $F(r_{+})=0$.
i.e.
\begin{eqnarray}
r_{+}^{6}+(3g^2-4m^2)r_{+}^{4}+3g^4r_{+}^2+g^6 &=& 0 ~.\label{horbd}
\end{eqnarray}

The largest real positive root of the equation gives the event horizon $(r_{+})$  of the
Bardeen black hole is given by

\begin{eqnarray}
 r_{+} &=& \frac{m}{\sqrt{3}} \frac{\sqrt{\left[16-24\left(\frac{g}{m}\right)^{2}+\{4-3\left(\frac{g}{m}\right)^{2} \}x^{\frac{1}{3}}+x^{\frac{2}{3}}\right]}}{x^{\frac{1}{6}}} \nonumber \\
  ~.\label{hobdr}
\end{eqnarray}
where,
$$
 x = 64-144 \left(\frac{g}{m}\right)^{2} +54\left(\frac{g}{m}\right)^{4} +
$$
\begin{equation}
6\sqrt{3}\left(\frac{g}{m}\right)^{3} \sqrt{27\left(\frac{g}{m}\right)^{2}-16} ~.\label{hobdr1}
\end{equation}

In the limit $g\rightarrow 0$, we recover  the horizon of the Schwarzschild
black hole i.e. $r_{+}=2m$.

The Bardeen space-time represents a regular black hole when $27g^2 \leq 16m^2$.
When $27g^2 < 16 m^2$, there are two horizons in the Bardeen
space-time, we may call it non-extremal Bardeen space-time
as in the non-extremal RN  space-time.
When $27g^2 = 16 m^2$, the two horizons are coincident, which correspond to
an extreme Bardeen black hole as in the RN black hole.

To derive the complete geodesic structure of the Bardeen black hole we shall follow
the pioneer book of S. Chandrasekhar \cite{sch} and J. B. Hartle \cite{hart}. To compute
the geodesic motion of the test particle in the equatorial plane we set $\dot{\theta}=0$
and $\theta=constant=\frac{\pi}{2}$. Since the space-time admits two Killing vectors
namely, $\zeta \equiv \partial_t$ and $\chi \equiv \partial_{\phi}$. Therefore
the quantities $E=-\zeta \cdot {\bf u}$  and  $L \equiv \chi \cdot {\bf u}$
are conserved along the geodesics, ${\bf u}$ is the four velocity
of the particle. Where $E$ and $L$ can be interpreted as conserved energy
and conserved angular momentum per unit mass respectively.

Thus in this coordinate chart, $E$ can be written as
\begin{eqnarray}
E &=&-\zeta \cdot {\bf u}={\cal F}(r)\,u^{t}~.\label{ptsBd}
\end{eqnarray}
and,
$L$ can be expressed as in terms of the metric
\begin{eqnarray}
r^2 \,u^{\phi}=L ~.\label{ppsBd}
\end{eqnarray}

From the normalization condition of the four velocity for
massive particles we find
\begin{eqnarray}
 {\bf u}^2 &=& \sigma ~.\label{norBd}
\end{eqnarray}
where $\sigma=-1$ for time-like geodesics, $\sigma=0$ for light-like geodesics
and $\sigma=+1$ for space-like geodesics.

Solving (\ref{ptsBd}) and (\ref{ppsBd}) for $u^{t}$ and $u^{\phi}$ we get
\begin{eqnarray}
u^{t} &=& \frac{E}{{\cal F}(r)} ~.\label{tdotBd}\\
u^{\phi} &=& \frac{L}{r^2}~.\label{phiBd}
\end{eqnarray}
where $E$ and $L$ are the energy and angular momentum per unit mass of the test particle.
Substituting these equations in (\ref{tdotBd}) and (\ref{phiBd}) in (\ref{norBd}), we
obtain the radial equation for the Bardeen space-time:
\begin{eqnarray}
(u^{r})^{2}=E^{2}-{\cal V}_{eff}=E^{2}-\left(\frac{L^{2}}{r^2}-\sigma \right){\cal F}(r)~.\label{radialbd}
\end{eqnarray}
where the standard effective potential for the geodesic motion of the
Bardeen space-time is given by
\begin{eqnarray}
{\cal V}_{eff}=\left(\frac{L^{2}}{r^2}-\sigma \right)\left(1-\frac{2mr^2}{(r^2+g^2)^{\frac{3}{2}}}\right) ~.\label{vrnBd}
\end{eqnarray}

\subsubsection{Particle orbits:}

The effective potential for time-like geodesics can be obtained from the above
equation (\ref{radialbd}) by putting $\sigma =-1$ as
\begin{eqnarray}
{\cal V}_{eff} &=& \left(1+\frac{L^{2}}{r^2}\right)\left(1-\frac{2mr^2}{(r^2+g^2)^{\frac{3}{2}}}\right) ~.\label{vrntbd}
\end{eqnarray}

To derive the circular geodesic motion of the test particle in
Bardeen space-time, we must impose the condition $\dot{r}=0$ at $r=r_{0}$. Thus one gets
from equation (\ref{radialbd})
\begin{eqnarray}
{\cal V}_{eff} &=& E^{2} ~.\label{vbd}
\end{eqnarray}
and,
\begin{eqnarray}
\frac{d{\cal V}_{eff}}{dr} &=& 0 ~.\label{dvdrbd}
\end{eqnarray}

Thus one can obtain the energy and angular momentum per unit mass of the
test particle along the circular orbits are:
\begin{eqnarray}
E^{2}_{0} &=& \frac{\left[(r_{0}^2+g^2)^{\frac{3}{2}}-2mr_{0}^2\right]^{2}}
{\sqrt{r_{0}^2+g^2}\left[(r_{0}^2+g^2)^{\frac{5}{2}}-3mr_{0}^4\right]} ~.\label{enggbd}
\end{eqnarray}
and,
\begin{eqnarray}
L^{2}_{0} &=& \frac{mr_{0}^{4}\left(r_{0}^2-2g^2\right)}
{\left[(r_{0}^2+g^2)^{\frac{5}{2}}-3mr_{0}^4\right]}~ .\label{anggbd}
\end{eqnarray}
Circular motion of the test particle to be exists when both
energy and angular momentum are real and finite.

Thus we must have
\begin{eqnarray}
(r_{0}^2+g^2)^{\frac{5}{2}}-3mr_{0}^4 > 0 \,\, \mbox{and} \,\, r_{0}>\sqrt{2}g ~ .
\end{eqnarray}

Circular orbits do not exist for all values of $r$, so from Eq. (\ref{enggbd}) and
Eq. (\ref{anggbd}), we can see that the denominator would be real only when
$$
(r_{0}^2+g^2)^{\frac{5}{2}}-3mr_{0}^4  \geq 0 ~.
$$
or
$$
r_{0}^{10}+(5g^2-9m^2)r_{0}^{8}+10g^4r_{0}^6+10g^6r_{0}^4+5g^8r_{0}^2+g^{10} \geq  0
$$

The limiting case of equality gives a circular orbit with infinite energy
per unit rest mass i.e. a photon orbit. This photon orbit is the inner most
boundary of the circular orbit for massive particles.

One can obtain MBCO for Bardeen space-time can be written as :

\begin{eqnarray}
r_{0}^6+(9g^2-16m^2)r_{0}^{4}+24g^4r_{0}^2 +16g^6 = 0 ~.\label{psbd}
\end{eqnarray}

Let $r_{0}=r_{mb}$ be the solution of the equation which gives the
radius of MBCO.

From astrophysical point of view the most important class of orbits
are ISCOs which can be derived from the second derivative of the
effective potential of time-like case.
i.e.
\begin{eqnarray}
\frac{d^2{\cal V}_{eff}}{dr^2} &=& 0 \label{pinbd}
\end{eqnarray}
Thus one may get the ISCO equation for the Bardeen space-time reads as
$$
r_{0}^{14}+(19g^2-36m^2)r_{0}^{12}+99g^4r_{0}^{10}+65g^6r_{0}^{8}-
$$
\begin{eqnarray}
160g^8r_{0}^6-144g^{10}r_{0}^{4}+64 g^{12}r_{0}^{2}+64g^{14} &=& 0 ~.\label{iscobd}
\end{eqnarray}

Let $r_{0}=r_{ISCO}$ be the real solution of the equation (\ref{iscobd}) which gives
the radius of the ISCO of Bardeen space-time.

In the limit $g \rightarrow 0$, we obtain the radius of ISCO for Schwarzschild
black hole which is $r_{ISCO}=6m$.

\subsubsection{Photon orbits:}

For null circular geodesics, the effective potential becomes
\begin{eqnarray}
{\cal U}_{eff} &=& \frac{L^2}{r^2}{\cal F}(r)
=\frac{L^2}{r^2} \left(1-\frac{2mr^2}{(r^2+g^2)^{\frac{3}{2}}}\right)
\end{eqnarray}
For circular null geodesics at $r=r_{c}$, we find
\begin{eqnarray}
 {\cal U}_{eff} &=& E^2
\end{eqnarray}
and,
\begin{eqnarray}
 \frac{d{\cal U}_{eff}}{dr} &=& 0
\end{eqnarray}

Thus one may  obtain the ratio of energy and angular momentum of the test particle
evaluated at $r=r_{c}$ for circular photon orbits are:
\begin{eqnarray}
 \frac{E_{c}}{L_{c}} &=& \pm \sqrt{\frac{(r_{c}^{2}+g^2)^{\frac{3}{2}}-2mr_{c}^2}{r_{c}^2(r_{c}^{2}
 +g^2)^{\frac{3}{2}}}}
\end{eqnarray}
and,
$$
r_{c}^{10}+(5g^2-9m^2)r_{c}^{8}+10g^{4}r_{c}^{6}+
$$
\begin{eqnarray}
10g^6r_{c}^{4}+5g^8r_{c}^2+g^{10} &=& 0 .~\label{ph1bd}
\end{eqnarray}

Let $r_{c}=r_{ph}$ be the  solution of the equation (\ref{ph1bd}) which gives
the radius of the CPO  of the  Bardeen space-time.
In the limit $g \rightarrow 0$, we obtain the radius of photon orbit for
Schwarzschild black hole which is $r_{ph}=3m$.

Let $D_{c}=\frac{L_{c}}{E_{c}}$ be the impact parameter for null circular geodesics
then
\begin{eqnarray}
 \frac{1}{D_{c}} &=& \frac{E_{c}}{L_{c}}=\sqrt{\frac{(r_{c}^{2}+g^2)^{\frac{3}{2}}-2mr_{c}^2}{r_{c}^2(r_{c}^{2}
 +g^2)^{\frac{3}{2}}}}
\end{eqnarray}
In the limit $g \rightarrow 0$, we obtain the impact parameter of the CPO
for the Schwarzschild black hole which is $D_{c}=3\sqrt{3}m$.

\subsection{CM energy and Particle collision:}

Now let us compute the CM energy  for the collision of two neutral
particles of same rest mass $m_{0}$ but different energy  coming from infinity
with $\frac{E_{1}}{m_{0}}=\frac{E_{2}}{m_{0}}=1$ and approaching the event horizon
(infinite red-shift surface) of the Bardeen black hole with different angular
momenta $L_{1}$ and $L_{2}$. Since our background is curved, so we need to define
the CM frame properly. BSW \cite{bsw} have been first derived the simple formula
which is valid in both flat and curved spacetime:
\begin{eqnarray}
\left(\frac{E_{cm}}{\sqrt{2}m_{0}}\right)^{2} &=&  1-g_{\mu\nu}u^{\mu}_{(1)}u^{\nu}_{(2)}~.\label{cm}
\end{eqnarray}
where $u^{\mu}_{(1)}$ and $u^{\mu}_{(2)}$ are the four velocity of the particles,
properly normalized by ${\bf u}.{\bf u}=-1$ (we have used the signature in the meric is (-+++)). This formula is of course well known in special relativity and in general relativity to ensure its validity.

We also assume throughout this work the geodesic motion of the colliding particles
confined in the equatorial plane. As we have previously said that the Bardeen space-time
admits a time-like isometry followed by the time-like Killing vector field $\zeta$ whose projection along the four velocity $\bf u$ of geodesics $\zeta.{\bf u}=-E$, is conserved along such geodesics. Similarly there is also the `angular momentum' $L=\chi.\bf u$ is conserved due to the rotational symmetry(where $\chi\equiv \partial_{\phi})$.
For time-like particles,  the components of the four velocity are
\begin{eqnarray}
  u^{t} &=& \dot{t}=\frac{{E}}{{\cal F}(r)}  \\
  u^{r} &=& \dot{r}=\pm \sqrt{E^{2}-{\cal F}(r)\left(1+\frac{L^{2}}{r^2}\right)} \label{eff}\\
  u^{\theta} &=& \dot{\theta} =0 \\
  u^{\phi} &=& \dot{\phi}=\frac{L}{r^2} ~.\label{utur}
\end{eqnarray}
and,
\begin{eqnarray}
u^{\mu}_{(1)} &=& \left(\frac{E_{1}}{{\cal F}(r)},~ -X_{1},~ 0,~\frac{L_{1}}{r^{2}}\right) ~.\label{u1}\\
u^{\mu}_{(2)} &=& \left(\frac{E_{2}}{{\cal F}(r)},~ -X_{2},~ 0,~\frac{L_{2}}{r^{2}}\right) ~.\label{u2} \\
\mbox{where}  \nonumber \\
X_{1} &=& \sqrt{E_{1}^{2}-{\cal F}(r)\left(1+\frac{L_{1}^{2}}{r^2}\right)} \\
X_{2} &=& \sqrt{E_{2}^{2}-{\cal F}(r)\left(1+\frac{L_{2}^{2}}{r^2}\right)}
\end{eqnarray}

\begin{figure}[t]
\begin{center}
{\includegraphics[width=0.45\textwidth]{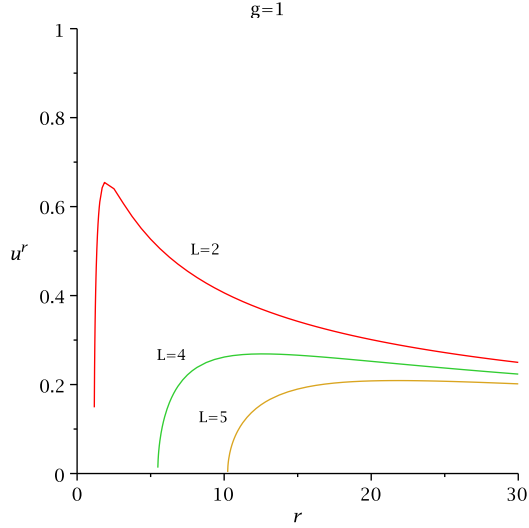}}
\end{center}
\caption{ The figure shows the variation  of $\dot{r}$  with $r$ for Bardeen 
black hole. Here, $m=1$. \label{rvbd}}
\end{figure}

\begin{figure}[t]
\begin{center}
{\includegraphics[width=0.45\textwidth]{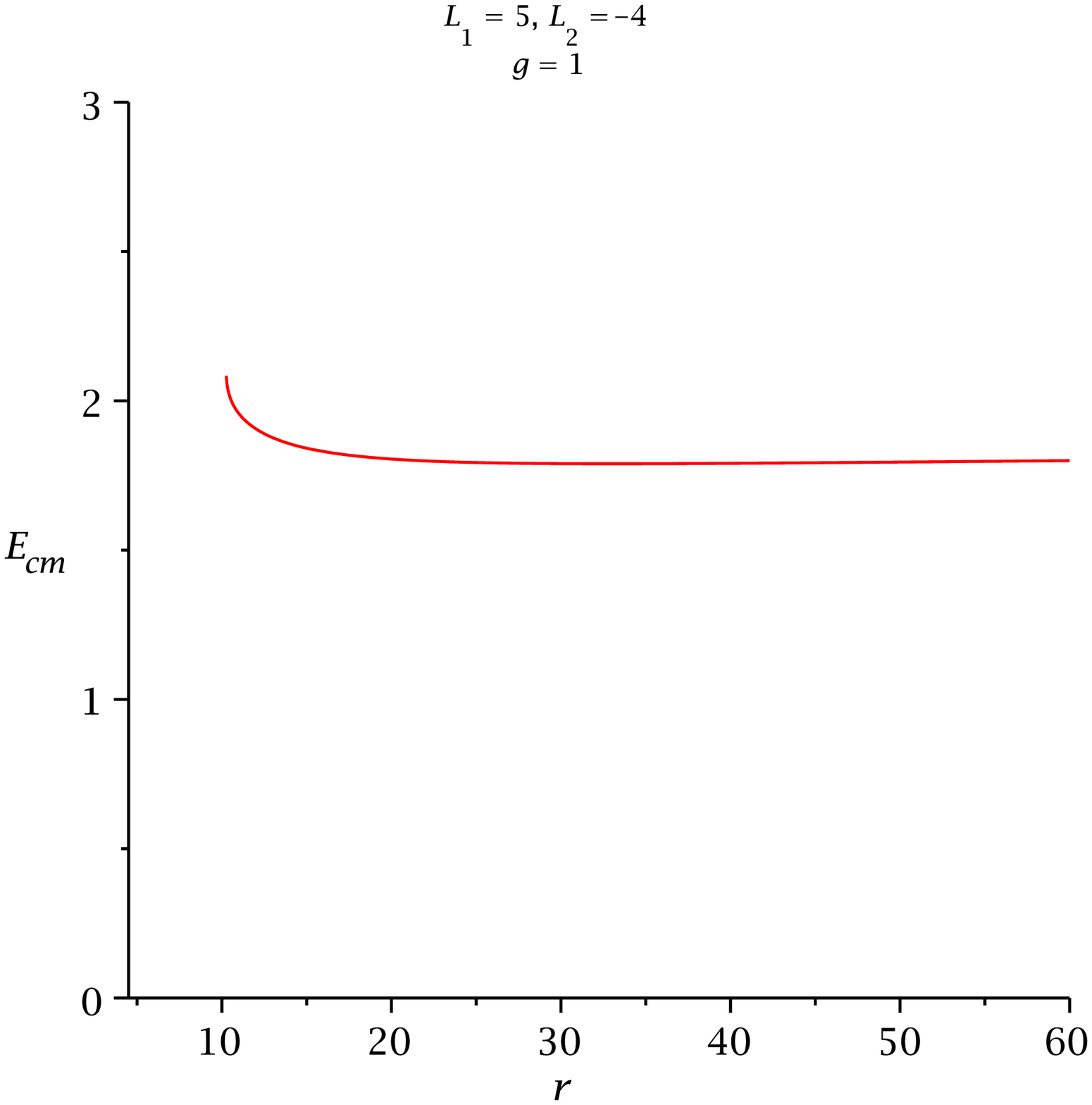}}
{\includegraphics[width=0.45\textwidth]{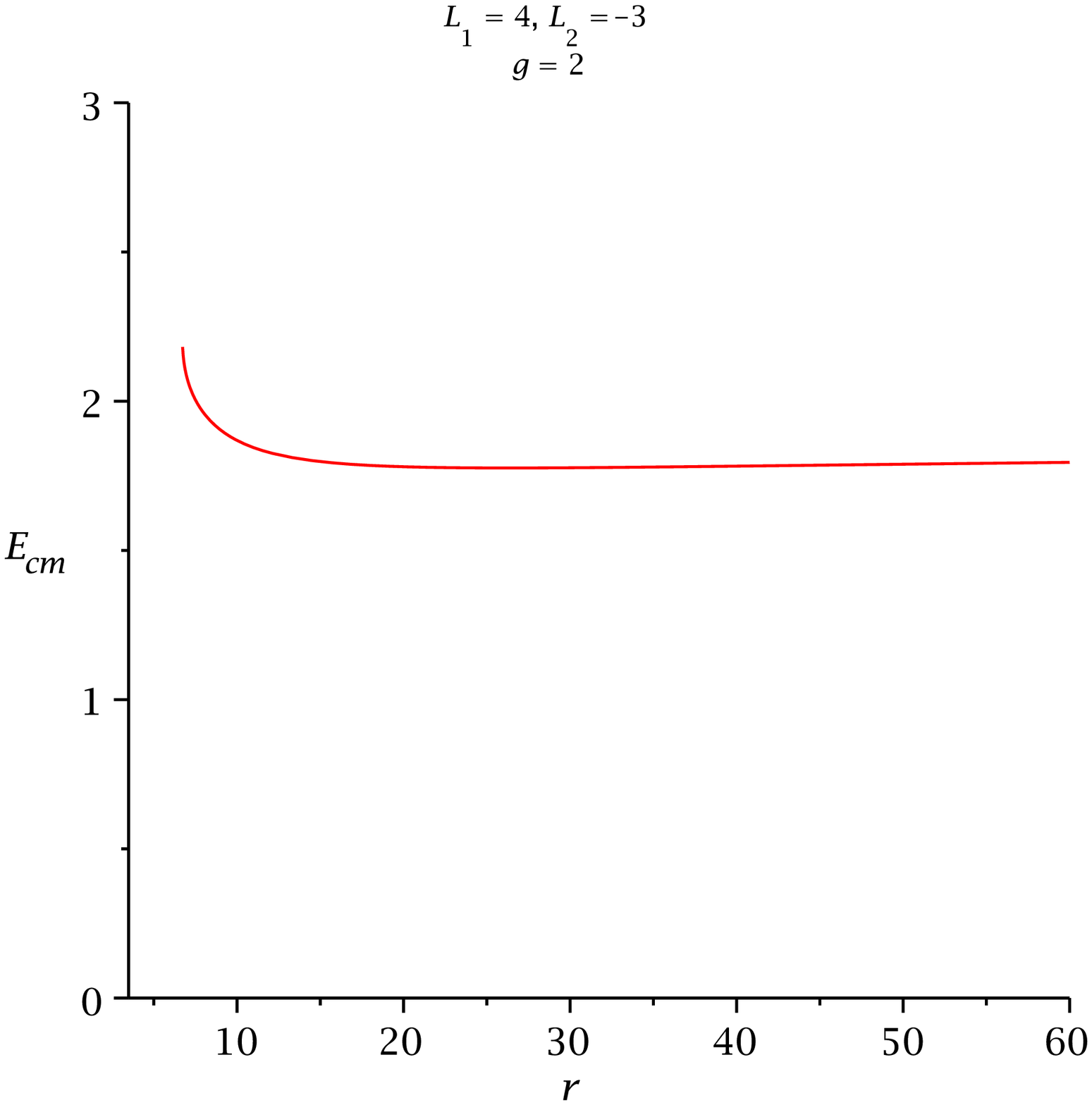}}
\end{center}
\caption{The figure shows the variation  of $E_{cm}$  with $r$ for Bardeen black hole. \label{cmbd}}
\end{figure}

\begin{figure}[t]
\begin{center}
{\includegraphics[width=0.45\textwidth]{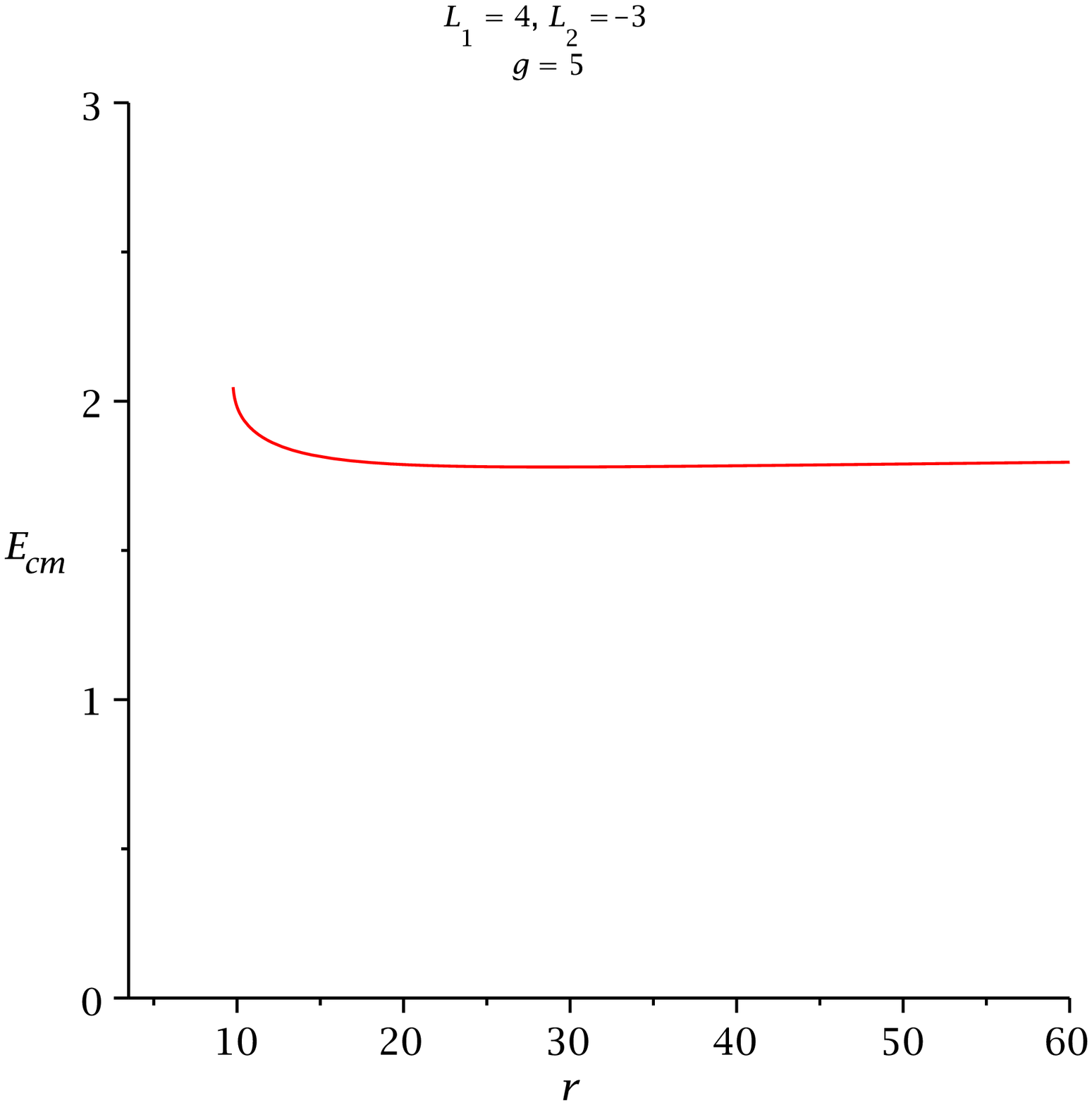}}
\end{center}
\caption{The figure shows the variation  of $E_{cm}$  with $r$ for Bardeen black hole
with different values of angular momentum. Here $g=5$.\label{cmbd2}}
\end{figure}

Substituting this in (\ref{cm}), we get the CM energy:
\begin{eqnarray}
\left(\frac{E_{cm}}{\sqrt{2}m_{0}}\right)^{2} &=&  1 +\frac{E_{1}E_{2}}{{\cal F}(r)}
-\frac{X_{1}X_{2}}{{\cal F}(r)}-\frac{L_{1}L_{2}}{r^{2}} ~.\label{cm1}
\end{eqnarray}
The figures[\ref{rvbd},\ref{cmbd},\ref{cmbd2}] depict the variation of radial velocity with $r$  and the variation of CM energy with $r$ for Bardeen black hole.

For simplicity, $E_{1}=E_{2}=1$ and substituting the value of ${\cal F}(r)$, we obtain
the CM energy near the event horizon ($r_{+}$) of the Bardeen space-time:

\begin{eqnarray}
E_{cm}\mid_{r\rightarrow r_{+}} &=& \sqrt{2}m_{0}\sqrt{\frac{4r_{+}^2+(L_{1}-L_{2})^{2}}{2r_{+}^2}} ~.\label{cmBard}
\end{eqnarray}
where $r_{+}$ is described in equation (\ref{hobdr}).

When we set $r_{+}=2m$,  we recover  the CM energy of the Schwarzschild black hole [\cite{bsw}]:
\begin{eqnarray}
E_{cm} &=& \sqrt{2}m_{0}\sqrt{\frac{16m^2+(L_{1}-L_{2})^{2}}{8m^2}} ~.\label{cmsch}
\end{eqnarray}
It is known that the maximum CM energy of the Schwarzschild
black hole occurs for the critical values of angular momentum parameter
i.e. $L_{1}=4m$ and $L_{2}=-4m$. Its value near the horizon is
$2\sqrt{5}m_{0}$ \cite{bau}.

The angular velocity of the Bardeen space-time at the $r_{+}$ is given by
\begin{eqnarray}
\Omega_{H}=\frac{\dot{\phi}}{\dot{t}}=\sqrt{\frac{m(r_{0}^2-2g^2)}{(r_{0}^2+g^2)^{\frac{5}{2}}}} ~.\label{omgBd}
\end{eqnarray}
The critical angular momenta $L_{i}$ can be written as

\begin{eqnarray}
L_{i} &=& \frac{E_{i}}{\Omega_{H}} ~.\label{li}
\end{eqnarray}
In the extremal cases, when $27g^2=16m^2$, the horizon is at $r_{0}=1.08m$ and it has been
already mentioned in \cite{bsw}, a new phenomenon would appear if one of the
particles participating in the collision has the critical angular momentum. If one
of the particles have the diverging angular momentum at the horizon i.e.
\begin{eqnarray}
L_{1}\mid_{r_{0}=1.08M} \rightarrow \infty ~.\label{liha}
\end{eqnarray}
Then for extremal Bardeen space-time, we get the infinite amounts of CM energy, i.e.
\begin{eqnarray}
 E_{cm} & \longmapsto &  \infty   ~.\label{cmdiv}
\end{eqnarray}

\section{CM energy of the collision near the horizon of the Ay\'{o}n-Beato and Garc\'{i}a  Space-time:}

In this section, we will investigate the CM energy of collision for two neutral particles
falling freely from rest at infinity in the horizon of a ABG black hole. This space-time
is also  a regular black hole space-time and singularity free solutions of the coupled
system of a non-linear electrodynamics and general relativity. The source is a nonlinear
electrodynamic field satisfying the WEC, which in the limit of weak field becomes the Maxwell field. We find the CM energy for this space-time can be infinitely high
when the black hole is only \emph{extremal}. Before computing the CM energy we shall
demonstrate shortly the geodesic structure of the ABG  space-time.

\subsection{Equatorial circular orbit in the ABG space-time:}

The metric of the ABG  space-time \cite{abg1,abg2,abg3,abg4,eva} is
given by
\begin{eqnarray}
ds^2=-{\cal G}(r)dt^{2}+\frac{dr^{2}}{{\cal G}(r)}
+r^{2}\left(d\theta^{2}+\sin^{2}\theta d\phi^{2}\right) ~.\label{abgs}
\end{eqnarray}
where the function ${\cal G}(r)$ is defined by
\begin{eqnarray}
{\cal G}(r) &=& 1- \frac{2mr^2}{(r^2+q^2)^{\frac{3}{2}}}+\frac{q^2r^2}{(r^2+q^2)^2}~.\label{aasym}
\end{eqnarray}
where $m$ is the mass of the black hole and $q$ is the monopole charge. The strength
of the radial electric field $E_{r}$ is given  by
\begin{eqnarray}
E_{r} &=& qr^{4}\left(\frac{r^{2}-5q^{2}}{(r^{2}+q^{2})^{4}}+\frac{15}{2}\frac{m}{(r^{2}+q^{2})^{7/2}}\right)
~.\label{radef}
\end{eqnarray}
We can see the behaviour of the function ${\cal G}(r)$ graphically.

\begin{figure}[t]
\begin{center}
{\includegraphics[width=0.45\textwidth]{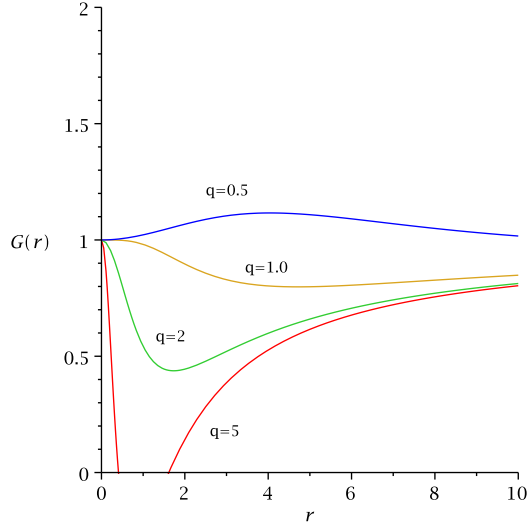}}
\end{center}
\caption{The figure shows the variation  of ${\cal G}(r)$  with $r$ for
different values of $q$.  Here, $m=1$. \label{abf}}
\end{figure}


This is also a first regular black hole solution in general relativity. The
source is a nonlinear electrodynamic field satisfying the WEC, which in the
limit of weak field becomes the Maxwell field.

It may also be noted that this metric function  asymptotically behaves as the
RN space-time\cite{abg1} i.e.,
\begin{eqnarray}
{\cal G}(r) \sim 1-\frac{2m}{r}+\frac{q^2}{r^2}+ O(\frac{1}{r^{3}}), \nonumber\\
E_{r} \sim \frac{q}{r^{2}}+O(\frac{1}{r^{3}}) ~.\label{asymabg}
\end{eqnarray}

The ABG black hole has an event horizon $(r_{+})$ which occur at $G(r_{+})=0$.
i.e.
\begin{eqnarray}
r_{+}^{8}+(6q^2-4m^2)r_{+}^{6}+(11q^4-4m^2q^2)r_{+}^{4} \nonumber \\
+6q^6r_{+}^2+q^8 = 0 ~.\label{horab}
\end{eqnarray}
In the limit $q\rightarrow 0$, we shall get the horizon of the Schwarzschild black hole i.e. $r_{+}=2m$.
The ABG space-time represents a regular black hole when $\mid q \mid \leq q_{c}$.
The value of $q_{c}$ is $q_{c}\approx 0.634m$.
When $\mid q \mid \leq q_{c}$, there are two horizons  in the ABG
space-time, we call it non-extremal ABG space-time as in the non-extremal
RN space-time.

When $\mid q \mid = q_{c}$, the two horizons are coincident at $r_{+} \sim 1.005 m $, which corresponds to an
extreme ABG black hole as in the RN black hole. The Carter-Penrose diagram of ABG space-time
is quite similar structure to the  RN black hole.

Proceeding analogously as in section 2, the radial equation that governs the
geodesic structure in the ABG space-time reads
\begin{eqnarray}
(u^{r})^{2}=\dot{r}^{2}=E^{2}-{\cal V}_{eff}=E^{2}-\left(\frac{L^{2}}{r^2}-\sigma \right){\cal G}(r)~.\label{radialabg}
\end{eqnarray}
where the effective potential for the geodesic motion of the  ABG space-time is given by

\begin{eqnarray}
{\cal V}_{eff}=\left(\frac{L^{2}}{r^2}-\sigma \right)\left(1- \frac{2mr^2}{(r^2+q^2)^{\frac{3}{2}}}+\frac{q^2r^2}{(r^2+q^2)^2}\right) ~.\label{vrnabg}
\end{eqnarray}

\subsubsection{Particle orbits:}

The effective potential for time-like geodesics can be written as using the
equation (\ref{radialabg}) by setting $\sigma =-1$ :
\begin{eqnarray}
{\cal V}_{eff} = \left(1+\frac{L^{2}}{r^2}\right)
\left(1- \frac{2mr^2}{(r^2+q^2)^{\frac{3}{2}}}+\frac{q^2r^2}{(r^2+q^2)^2}\right) ~.\label{vrntabg}
\end{eqnarray}

To derive the circular geodesic motion of the test particle in
ABG space-time, we must use the condition $\dot{r}=0$ at $r=r_{0}$.
From equation (\ref{radialabg}), one gets
\begin{eqnarray}
{\cal V}_{eff} &=& E^{2} ~.\label{vabg}
\end{eqnarray}
and
\begin{eqnarray}
\frac{d{\cal V}_{eff}}{dr} &=& 0 ~.\label{dvdrabg}
\end{eqnarray}

Thus one can obtain the energy and angular momentum per unit mass of the
test particle along the circular orbits :
\begin{eqnarray}
E^{2}_{0}=\frac{\left[(r_{0}^2+q^2)^{2}-2mr_{0}^2\sqrt{r_{0}^2+q^2}+q^2r_{0}^2\right]^{2}}
{(r_{0}^2+q^2)\left[(r_{0}^2+q^2)^{3}-3mr_{0}^{4}\sqrt{r_{0}^2+q^2}+2q^2r_{0}^4\right]} ~.\label{enggabg}
\end{eqnarray}
and,
\begin{eqnarray}
L^{2}_{0} = \frac{r_{0}^{4}\left[m(r_{0}^2-2q^2)\sqrt{r_{0}^2+q^2}-q^2(r_{0}^2-q^2)\right]}
{\left[(r_{0}^2+q^2)^{3}-3mr_{0}^{4}\sqrt{r_{0}^2+q^2}+2q^2r_{0}^4\right]}~ .\label{anggabg}
\end{eqnarray}
Circular motion of the test particle to be exists for ABG space-time
when both energy and angular momentum are real and finite.

Thus  we get the inequality:
$$
(r_{0}^2+q^2)^{3}-3mr_{0}^{4}\sqrt{r_{0}^2+q^2}+2q^2r_{0}^4 > 0
$$
and
$$
r_{0}>q\sqrt{\frac{2m\sqrt{r_{0}^2+q^2}-q^2}{m\sqrt{r_{0}^2+q^2}-q^2}}~.
$$

Circular orbits do not exist for all values of $r$, so from Eq. (\ref{enggabg}) and
Eq. (\ref{anggabg}), we can see that the denominator would be real only when

\begin{eqnarray}
(r_{0}^2+q^2)^{3}-3mr_{0}^{4}\sqrt{r_{0}^2+q^2}+2q^2r_{0}^4 &\geq&  0 ~ . \\
\mbox{or} \nonumber
\end{eqnarray}
$$
r_{0}^{12}+(10q^2-9m^2)r_{0}^{10}-(9m^2q^2-31q^4)r_{0}^{8} +
$$
\begin{eqnarray}
32q^6r_{0}^6+19q^8r_{0}^4+6q^{10}r_{0}^{2}+q^{12} &\geq &  0
\end{eqnarray}

The limiting case of equality indicates a circular orbit with diverging energy
per unit rest mass i.e. a photon orbit. This photon orbit is the inner most
boundary of the circular orbits for time-like particles.

The equation of MBCO for ABG space-time
looks like:
$$
m^2r_{0}^{10}-(16m^4-3m^2q^2)r_{0}^{8} +
$$
$$
(99m^2q^4-32m^4q^2) r_{0}^6 -(16m^4q^4-23m^2q^6-9q^8) r_{0}^{4} +
$$
\begin{eqnarray}
(72m^2q^8-12q^{10})r_{0}^{2}+(16m^2q^{10}-4q^{12}) &=& 0 ~.\label{psabg}
\end{eqnarray}
Let $r_{0}=r_{mb}$ be the solution of the equation which gives the
radius of MBCO close to the black hole.

The ISCO equation can be obtained from  the second derivative of the
effective potential of time-like case.
i.e.
\begin{eqnarray}
\frac{d^2{\cal V}_{eff}}{dr^2} &=& 0 \label{piebd}
\end{eqnarray}
Thus one may get the ISCO equation for the ABG space-time reads as
$$
m^2 r_{0}^{18}-(36m^4-39m^2q^2+4q^{4}) r_{0}^{16} +
$$
$$
(97m^2q^4-72m^4q^2+40q^{6}) r_{0}^{14}-(36m^4q^4-97m^2q^6+52q^8) r_{0}^{12}
$$
$$
- (89m^2q^8+216q^{10}) r_{0}^{10} - (357m^2q^{10}+272q^{12}) r_{0}^{8} -
$$
$$
(292m^2q^{12}+104q^{14}) r_{0}^{6}+(16m^2q^{14}+12q^{16}) r_{0}^{4} +
$$
\begin{eqnarray}
(144m^2q^{16}+24q^{18})r_{0}^{2} + 4q^{18}(16m^2-q^2) &=& 0 ~.\label{iscoabg}
\end{eqnarray}
Let $r_{0}=r_{ISCO}$ be the real solution of the equation (\ref{iscoabg}) which gives
the radius of the ISCO of ABG space-time.

In the limit $q \rightarrow 0$, we obtain the radius of ISCO for Schwarzschild
black hole which is $r_{ISCO}=6m$.

\subsubsection{Photon orbits:}

For null circular geodesics, the effective potential becomes
\begin{eqnarray}
{\cal U}_{eff} &=& \frac{L^2}{r^2}{\cal G}(r) \nonumber \\[4mm]
&=& \frac{L^2}{r^2}\left(1-\frac{2mr^2}{(r^2+q^2)^{\frac{3}{2}}}+\frac{q^2r^2}{(r^2+q^2)^2}\right)
\end{eqnarray}

For circular null geodesics at $r=r_{c}$, we find
\begin{eqnarray}
 {\cal U}_{eff} &=& E^2
\end{eqnarray}
and,
\begin{eqnarray}
 \frac{d{\cal U}_{eff}}{dr} &=& 0
\end{eqnarray}

Thus one may  obtain the ratio of energy and angular momentum of the test particle
evaluated at $r=r_{c}$ for CPO is

\begin{eqnarray}
 \frac{E_{c}}{L_{c}} &=& \pm \sqrt{\frac{1}{r_{c}^2}\left(1- \frac{2mr_{c}^2}{(r_{c}^2+q^2)^{\frac{3}{2}}}+\frac{q^2r_{c}^2}{(r_{c}^2+q^2)^2}\right)}
\end{eqnarray}
and,
$$
r_{c}^{12}+(10q^2-9m^2)r_{c}^{10}-(9m^2q^2-31q^4)r_{c}^{8} +
$$
\begin{eqnarray}
32q^6r_{c}^6+19q^8r_{c}^4 + 6q^{10}r_{c}^{2}+q^{12} &=& 0 .~\label{ph1abg}
\end{eqnarray}
Let $D_{c}=\frac{L_{c}}{E_{c}}$ be the impact parameter for null circular geodesics
then
\begin{eqnarray}
 \frac{1}{D_{c}} = \frac{E_{c}}{L_{c}}=\sqrt{\frac{1}{r_{c}^2}\left(1- \frac{2mr_{c}^2}{(r_{c}^2+q^2)^{\frac{3}{2}}}+\frac{q^2r_{c}^2}{(r_{c}^2+q^2)^2}\right)}
\end{eqnarray}
Let $r_{c}=r_{ph}$ be the  solution of the equation (\ref{ph1abg}) which gives
the radius of the photon orbit  of the  ABG space-time. In the limit $q \rightarrow 0$, we recover the
 CPO of Schwarzschild black hole which is $r_{ph}=3m$.

\subsection{CM Energy for ABG space-time:}

Now let us compute the CM energy  for the collision of two neutral
particles coming from infinity with $\frac{E_{1}}{m_{0}}=\frac{E_{2}}{m_{0}}=1$ and
approaching the ABG space-time with different angular momenta $L_{1}$ and $L_{2}$.

Since the ABG space-time has also Killing symmetries followed by the Killing vector field thus energy $(E)$ and angular momentum ($L$)  are conserved quantities as we have defined in case of Bardeen space-time.

Therefore for massive particles of ABG space-time, the components of the four velocity are
\begin{eqnarray}
  u^{t} &=& \frac{{E}}{{\cal G}(r)}  \\
  u^{r} &=& \pm \sqrt{E^{2}-{\cal G}(r)\left(1+\frac{L^{2}}{r^2}\right)} \label{efAB}\\
  u^{\theta} &=& 0 \\
  u^{\phi} &=& \frac{L}{r^2} ~.\label{uturAB}
\end{eqnarray}
and,
\begin{eqnarray}
u^{\mu}_{(1)} &=& \left( \frac{E_{1}}{{\cal G}(r)},~ -Y_{1} ,~ 0,~\frac{L_{1}}{r^{2}}\right) ~.\label{u1AB}\\
u^{\mu}_{(2)} &=& \left( \frac{E_{2}}{{\cal G}(r)},~ -Y_{2} ,~ 0,~\frac{L_{2}}{r^{2}}\right) ~.\label{u2AB}\\
\mbox{where} \nonumber \\
Y_{1} &=& \sqrt{E_{1}^{2}-{\cal G}(r)\left(1+\frac{L_{1}^{2}}{r^2}\right)}\\
Y_{2} &=& \sqrt{E_{2}^{2}-{\cal G}(r)\left(1+\frac{L_{2}^{2}}{r^2}\right)}
\end{eqnarray}
We can see graphically the variation of $\dot{r}$ with $r$ and the variation
of CM energy with $r$ for ABG space-time.

\begin{figure}[t]
\begin{center}
{\includegraphics[width=0.45\textwidth]{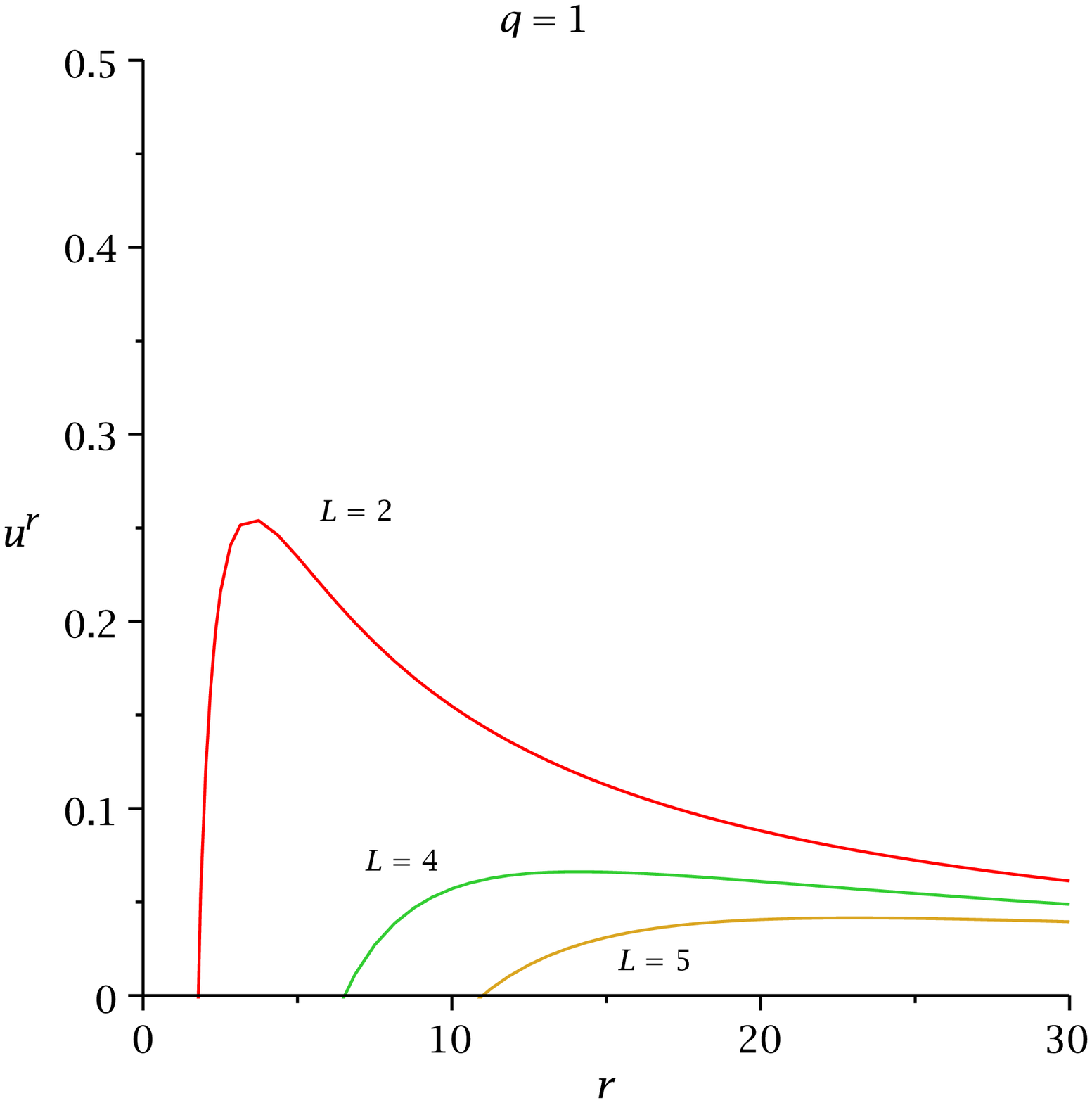}}
\end{center}
\caption{The figure shows the variation  of $\dot{r}$  with $r$ for ABG space-time. 
Here, $m=1, q=1$. \label{abgv1}}
\end{figure}


\begin{figure}[t]
\begin{center}
{\includegraphics[width=0.45\textwidth]{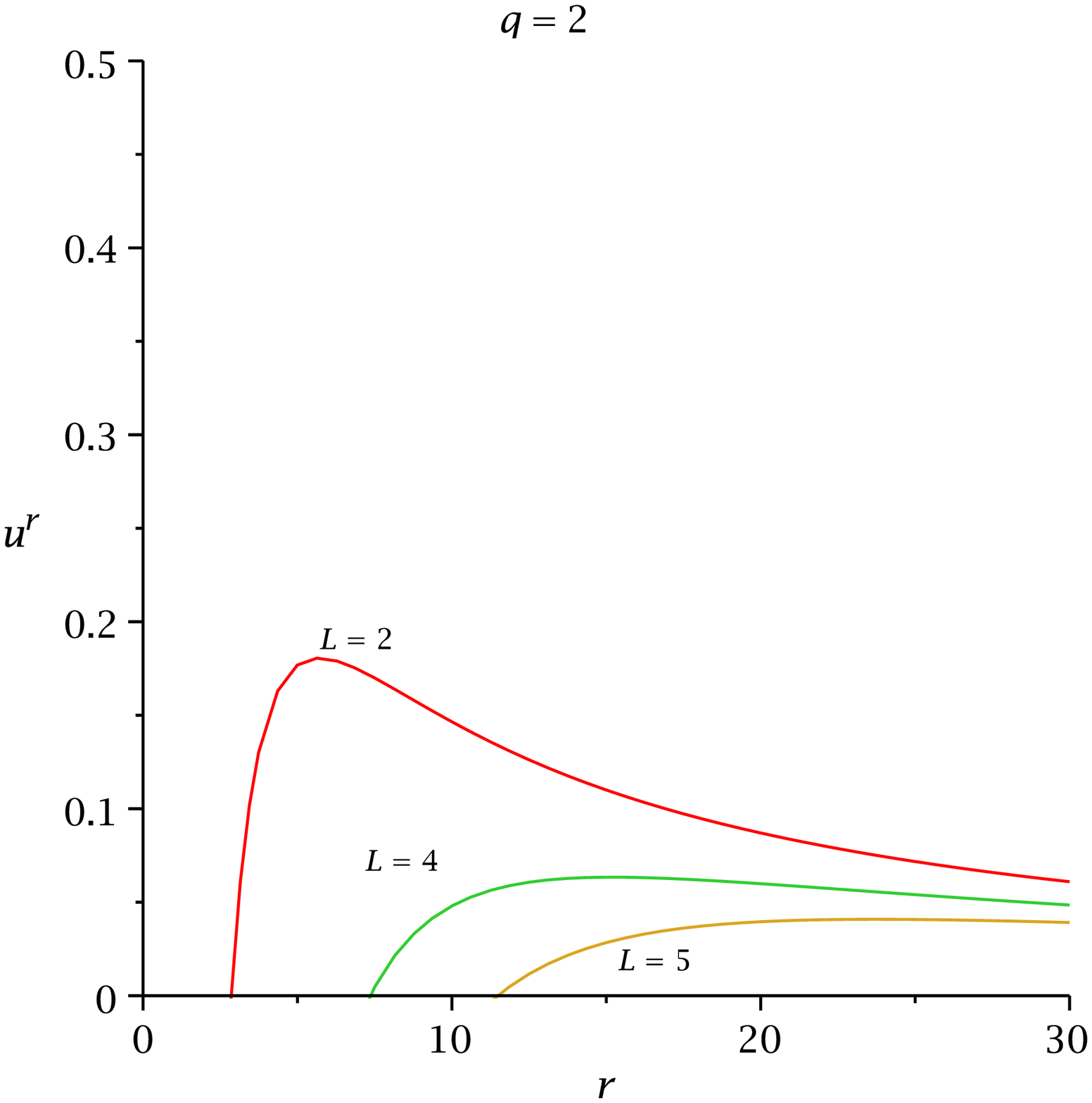}}
{\includegraphics[width=0.45\textwidth]{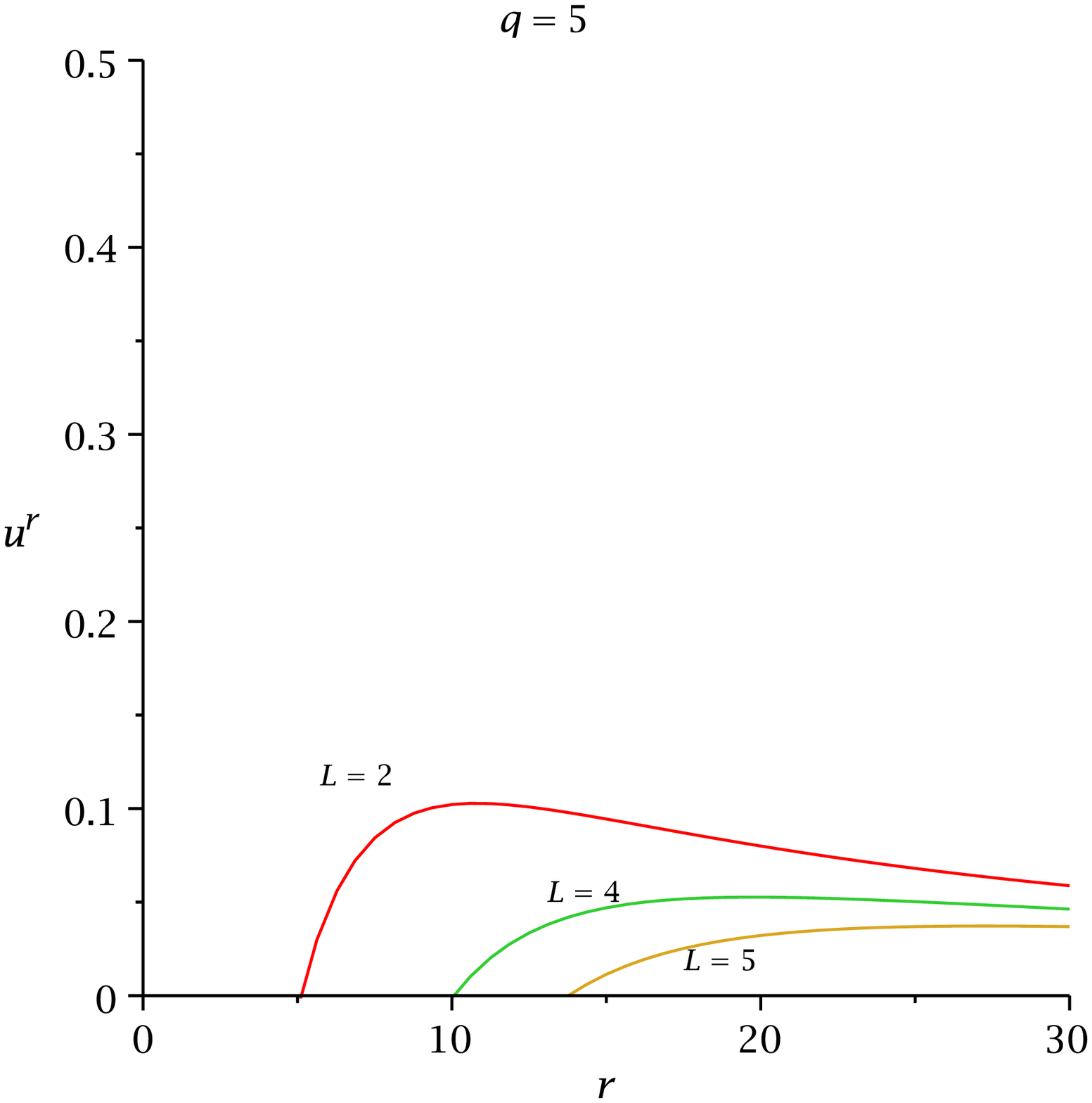}}
\end{center}
\caption{The figure shows the variation  of $\dot{r}$  with $r$ for ABG space-time.
Here, $m=1$. \label{abgv2}}
\end{figure}



\begin{figure}[t]
\begin{center}
{\includegraphics[width=0.45\textwidth]{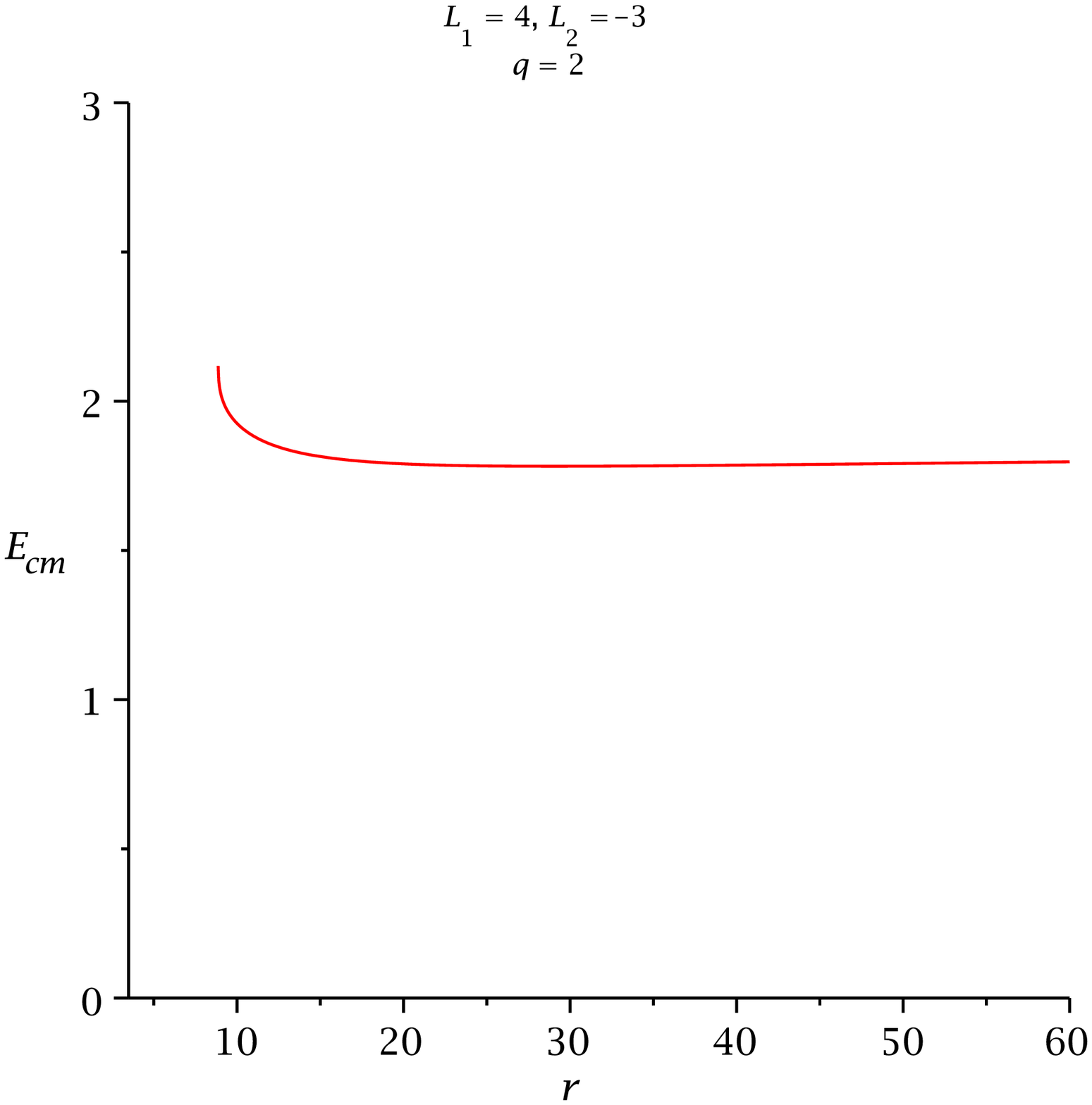}}
{\includegraphics[width=0.45\textwidth]{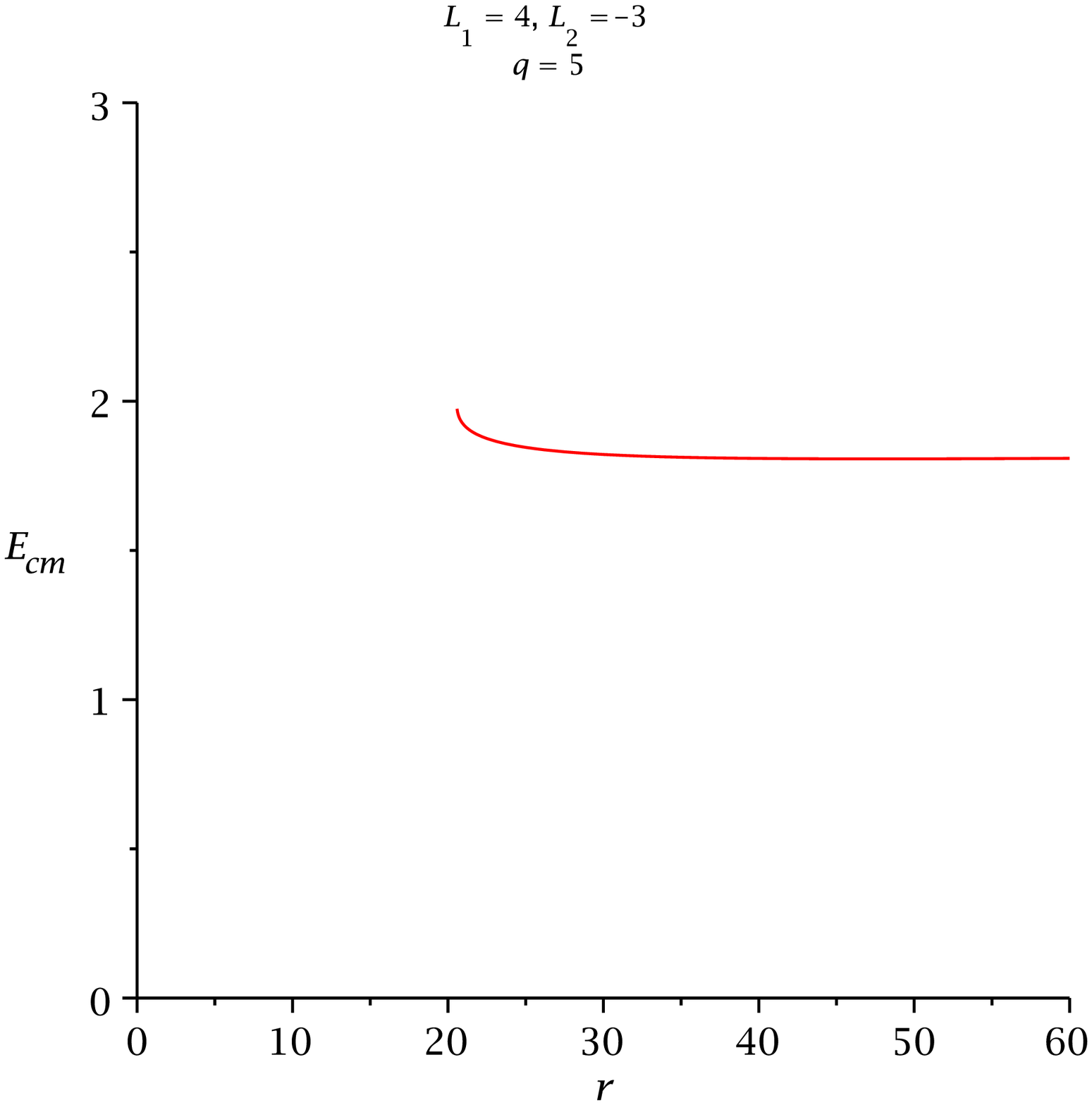}}
\end{center}
\caption{The figure shows the variation  of $E_{cm}$  with $r$ for ABG black hole
with different values of angular momentum. \label{abgc1}}
\end{figure}



Substituting this in (\ref{cm}), we get the center of mass energy for ABG space-time:
\begin{eqnarray}
\left(\frac{E_{cm}}{\sqrt{2}m_{0}}\right)^{2} &=&  1 +\frac{E_{1}E_{2}}{{\cal G}(r)}-
\frac{Y_{1}Y_{2}}{{\cal G}(r)}-\frac{L_{1}L_{2}}{r^2} ~.\label{cmAB}
\end{eqnarray}

For simplicity, $E_{1}=E_{2}=1$ and putting the value of ${\cal G}(r)$, we obtain
the CM energy near the event horizon ($r_{+}$) of the ABG space-time:
\begin{eqnarray}
E_{cm}\mid_{r\rightarrow r_{+}} &=& \sqrt{2}m_{0}\sqrt{\frac{4r_{+}^2+(L_{1}-L_{2})^{2}}{2r_{+}^2}} ~.\label{cmabg}
\end{eqnarray}
where $r_{+}$ is the root of the  given Eq.  in (\ref{horab}) .

When we set $r_{+}=2m$,  we get  the CM energy of the Schwarzschild black hole:
\begin{eqnarray}
E_{cm} &=& \sqrt{2}m_{0}\sqrt{\frac{16m^2+(L_{1}-L_{2})^{2}}{8m^2}} ~.\label{cmsch1}
\end{eqnarray}

The angular velocity of the ABG space-time at the  event horizon  $r_{+}$ is given by
\begin{eqnarray}
\Omega_{H}=\frac{\dot{\phi}}{\dot{t}}
=\sqrt{\frac{m(r_{0}^2-2q^2)}{(r_{0}^2+q^2)^{\frac{5}{2}}}-\frac{q^2(r_{0}^2-q^2)}{(r_{0}^2+q^2)^3}} ~.\label{omgabg}
\end{eqnarray}
The critical angular momenta $L_{i}$ may be written as

\begin{eqnarray}
L_{i} &=& \frac{E_{i}}{\Omega_{H}} ~.\label{lia}
\end{eqnarray}
At the extremal cases, when $q=q_{c}$, the horizon is at $r_{0}=1.005m$ and
if one of the values of critical angular momenta diverge i.e.
\begin{eqnarray}
L_{1}\mid_{r_{0}=1.005M} \rightarrow \infty ~.\label{li2}
\end{eqnarray}
Therefore for extremal ABG space-time, we obtain the infinite amount of
CM energy, i.e.
\begin{eqnarray}
 E_{cm} & \longmapsto &  \infty   ~.\label{cmdiv1}
\end{eqnarray}

\section{CM energy of the collision near the horizon of the Hayward  Black hole:}

Finally, in this section, we shall perform similar analysis for another
interesting regular black hole i.e. Hayward black hole which was
suggested by Hayward in 2006 for the process of a regular black hole
formation and evaporation.

The line element for the Hayward black hole \cite{hay} is given by
\begin{eqnarray}
ds^2=-{\cal H}(r)dt^{2}+\frac{dr^{2}}{{\cal H}(r)}+
r^{2}\left(d\theta^{2}+\sin^{2}\theta d\phi^{2} \right) ~.\label{hay}
\end{eqnarray}
where the function ${\cal H}(r)$ is defined by
\begin{eqnarray}
{\cal H}(r) &=& 1- \frac{2mr^2}{(r^3+2\alpha^2)}.\\
\mbox{and} \nonumber \\
\alpha^2 &=& ml^2.
\end{eqnarray}
where $m$ is the mass of the black hole and $l$ is a free parameter.
The behaviour of the metric function ${\cal H}(r)$ for Hayward black hole
for different values of $l$ can be seen from the following diagram.
\begin{figure}[t]
\begin{center}
{\includegraphics[width=0.45\textwidth]{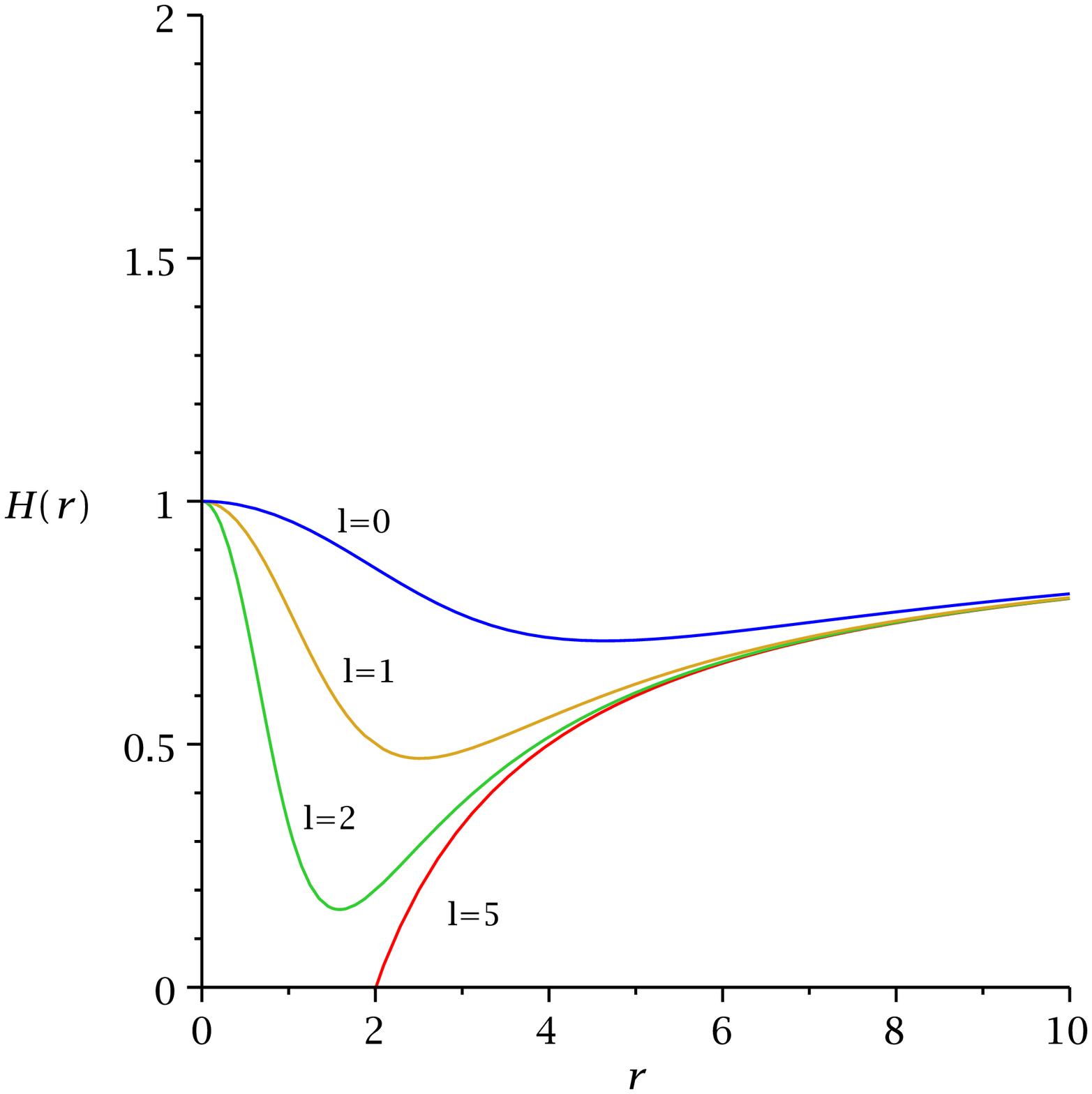}}
\end{center}
\caption{The figure shows the variation  of ${\cal H}(r)$  with $r$ for
different values of $l$.  Here, $m=1$. \label{hay1}}
\end{figure}


Hayward first use such types of  metric for  the formation and evaporation of a
non-singular black hole \cite{hay}. The metric describes a static, spherically
symmetric and asymptotically flat, have regular centers and for which the resulting
energy-momentum tensor satisfying the WEC. The Carter-Penrose  diagram is
similar to that of RN  space-time, with the internal
singularities replaced by regular centers.

Noticed that this metric function asymptotically i.e. at $r\rightarrow \infty$ behaves
as\cite{hay}
\begin{eqnarray}
{\cal H}(r) \sim 1-\frac{2m}{r} ~.\label{asymh}
\end{eqnarray}
Furthermore, if $r\rightarrow 0$, the metric behaves as the de-Sitter space-time:
\begin{eqnarray}
{\cal H}(r) \sim 1-\frac{r^2}{l^2} ~.\label{asymh1}
\end{eqnarray}

The Hayward black hole has an event horizon $(r_{+})$ which occur at $H(r_{+})=0$.
i.e.
\begin{eqnarray}
r_{+}^{3}-2mr_{+}^2+2ml^2 &=& 0 ~.\label{horHay}
\end{eqnarray}

The largest real positive root of the equation is given by
\begin{eqnarray}
 r_{+} &=& \frac{m}{3}\left[2+z^{\frac{1}{3}}+\frac{4}{z^{\frac{1}{3}}}\right] ~.\label{hoha1}
\end{eqnarray}
where
\begin{eqnarray}
z = 8-27 \left(\frac{l}{m}\right)^{2} +9\sqrt{3}\left(\frac{l}{m}\right)
\sqrt{27\left(\frac{l}{m}\right)^{2}-16} ~.\label{horHay2}
\end{eqnarray}
when $l=0$, we recover the Schwarzschild black hole horizon ($r_{+}=2m$).

The Hayward space-time represents a regular black hole when $27l^2 \leq 16m^2$.

When $27l^2 < 16 m^2$, there are two horizons in the Hayward
space-time, we call it non-extremal Hayward space-time.

When $27l^2 = 16 m^2$, the two horizons  merge, which correspond to
an extreme Hayward black hole.

To study the geodesic motion for this space-time we shall perform
similar analysis as we have done in section 2 using Killing
symmetries.

Thus  the radial equation that govern the geodesic motion in the
equatorial plane for the Hayward space-time can be written as:
\begin{eqnarray}
(u^{r})^{2}=E^{2}-{\cal V}_{eff}=E^{2}-\left(\frac{L^{2}}{r^2}-\sigma \right){\cal H}(r)~.\label{radialh}
\end{eqnarray}
where the standard effective potential that describe the geodesic motion of the
Hayward space-time is

\begin{eqnarray}
{\cal V}_{eff}=\left(\frac{L^{2}}{r^2}-\sigma \right)
\left(\frac{2mr^2}{r^3+2ml^2} \right) ~.\label{vrnh}
\end{eqnarray}

\subsubsection{Particle orbits:}

The effective potential for time-like geodesics for the Hayward space-time becomes

\begin{eqnarray}
{\cal V}_{eff} &=& \left(1+\frac{L^{2}}{r^2}\right)\left(1-\frac{2mr^2}{(r^3+2l^2m)}\right) ~.\label{vrnth}
\end{eqnarray}

To derive the circular geodesic motion of the test particle in the
Hayward space-time, we must have the condition $\dot{r}=0$ at
$r=r_{0}$. Thus one gets, from equation (\ref{radialh})
\begin{eqnarray}
{\cal V}_{eff} &=& E^{2} ~.\label{vh}
\end{eqnarray}
and,
\begin{eqnarray}
\frac{d{\cal V}_{eff}}{dr} &=& 0 ~.\label{dvdrh}
\end{eqnarray}

A straightforward calculation implies  that the energy and angular momentum
per unit mass of the test particle along the circular orbits are:

\begin{eqnarray}
E^{2}_{0} &=& \frac{\left[r_{0}^3+2ml^2-2mr_{0}^2\right]^{2}}
{\left[r_{0}^6-3mr_{0}^{5}+4ml^{2}r_{0}^3+4m^2l^{4}\right]} ~.\label{enggh}
\end{eqnarray}
and,
\begin{eqnarray}
L^{2}_{0} &=& \frac{mr_{0}^{4}\left(r_{0}^3-4ml^{2}\right)}
{\left[r_{0}^6-3mr_{0}^5+4ml^{2}r_{0}^3+4m^2l^{4}\right]}~ .\label{anggh}
\end{eqnarray}
The condition for circular motion to be exists in the Hayward space-time
when both energy ($E_{0}$) and angular momentum ($L_{0}$) are real and finite.

Thus we have the condition:
\begin{eqnarray}
r_{0}^6-3mr_{0}^5+4ml^{2}r_{0}^3+4m^2l^{4}> 0 \,\, \mbox{and}
\,\, r_{0}>(4ml^2)^{\frac{1}{3}} ~ .
\end{eqnarray}
Circular orbits do not exist for all radii, so from Eq. (\ref{enggh}) and
Eq. (\ref{anggh}), we can find that the denominator would be real only when

\begin{eqnarray}
r_{0}^6-3mr_{0}^5+4ml^{2}r_{0}^3+4m^2l^{4} \geq 0 ~ .
\end{eqnarray}

The limiting case of equality gives a circular orbit with infinite energy
per unit rest mass i.e. a photon orbit. This photon orbit is the inner most
boundary of the circular orbits for massive particles.

One can obtain MBCO for Hayward space-time
would be
\begin{eqnarray}
r_{0}^3-4mr_{0}^{2}+8ml^{2} &=& 0 ~.\label{psh}
\end{eqnarray}

Using MAPLE software, we can find the real positive root of the Eq.(\ref{psh})
which gives the radius of MBCO closest to the black hole is given by

\begin{eqnarray}
 r_{mb} &=& \frac{m}{3}\left[4+y^{\frac{1}{3}}+\frac{16}{y^{\frac{1}{3}}}\right] ~.\label{mbha1}
\end{eqnarray}
where,
$$
y = 64-108\left(\frac{l}{m}\right)^{2} +
$$
\begin{eqnarray}
12\sqrt{3}\left(\frac{l}{m}\right) \sqrt{27\left(\frac{l}{m}\right)^{2}-32}~.\label{horHay3}
\end{eqnarray}
From an astrophysical significance  the most important class of orbits
are ISCOs which can be calculated from the second derivative of the
effective potential of time-like case.
i.e.
\begin{eqnarray}
\frac{d^2{\cal V}_{eff}}{dr^2} &=& 0 \label{pibd}
\end{eqnarray}
Thus one would obtain the ISCO equation for the Hayward space-time:
$$
r_{0}^{9}-6mr_{0}^{8}+24ml^2r_{0}^{6} -
$$
\begin{eqnarray}
12m^{2}l^2r_{0}^5+12m^2l^{4}r_{0}^{3}-64 m^{3}l^{6} &=& 0 ~.\label{iscoh}
\end{eqnarray}

Let $r_{0}=r_{ISCO}$ be the real solution of the equation (\ref{iscoh}) which gives
the radius of the ISCO of Hayward space-time.

In the limit $l \rightarrow 0$, we obtain the radius of ISCO for Schwarzschild
black hole which is $r_{ISCO}=6m$.

\subsubsection{Photon orbits:}

For null circular geodesics, the effective potential becomes
\begin{eqnarray}
{\cal U}_{eff} &=& \frac{L^2}{r^2}{\cal H}(r)
=\frac{L^2}{r^2} \left(1-\frac{2mr^2}{(r^3+2ml^2)}\right)
\end{eqnarray}
For circular null geodesics at $r=r_{c}$, we find
\begin{eqnarray}
 {\cal U}_{eff} &=& E^2
\end{eqnarray}
and
\begin{eqnarray}
 \frac{d{\cal U}_{eff}}{dr} &=& 0
\end{eqnarray}

Thus one may  obtain for Hayward space-time, the ratio of energy and angular
momentum of the test particle evaluated at $r=r_{c}$ for circular photon orbits  are:

\begin{eqnarray}
 \frac{E_{c}}{L_{c}} &=& \pm \sqrt{\frac{1}{r_{c}^{2}}\left(1-\frac{2mr_{c}^2}{(r_{c}^3+2ml^2)}\right)}
\end{eqnarray}
and,
\begin{eqnarray}
r_{c}^6-3mr_{c}^5+4ml^{2}r_{c}^3+4m^2l^{4}&=& 0 .~\label{ph1h}
\end{eqnarray}
Let $D_{c}=\frac{L_{c}}{E_{c}}$ be the impact parameter for null circular geodesics
then
\begin{eqnarray}
\frac{1}{D_{c}} &=& \frac{E_{c}}{L_{c}}=\sqrt{\frac{(r_{c}^{3}+2ml^2)-2mr_{c}^2}{r_{c}^2(r_{c}^{3}
+2ml^2)}}
\end{eqnarray}
Let $r_{c}=r_{ph}$ be the  solution of the equation (\ref{ph1h}) which gives
the radius of the  circular photon orbit  of the  Hayward space-time.

\subsection{CM Energy for Hayward space-time:}

The components of the four velocity in terms of the energy and
angular momentum due to the Killing symmetries of the space-time for
time-like particles are

\begin{eqnarray}
  u^{t} &=& \frac{{E}}{{\cal H}(r)}  \\
  u^{r} &=& \pm \sqrt{E^{2}-{\cal H}(r)\left(1+\frac{L^{2}}{r^2}\right)} \label{efHS}\\
  u^{\theta} &=& 0 \\
  u^{\phi} &=& \frac{L}{r^2} ~.\label{uturHS}
\end{eqnarray}
and,
\begin{eqnarray}
u^{\mu}_{1} &=& \left( \frac{E_{1}}{{\cal H}(r)},~ -Z_{1} ,~ 0,~\frac{L_{1}}{r^{2}}\right) ~.\label{u1HS}\\
u^{\mu}_{2} &=& \left( \frac{E_{2}}{{\cal H}(r)},~ -Z_{2} ,~ 0,~\frac{L_{2}}{r^{2}}\right) ~.\label{u2HS}\\
\mbox{where} \nonumber \\
Z_{1} &=& \sqrt{E_{1}^{2}-{\cal H}(r)\left(1+\frac{L_{1}^{2}}{r^2}\right)}\\
Z_{2} &=& \sqrt{E_{2}^{2}-{\cal H}(r)\left(1+\frac{L_{2}^{2}}{r^2}\right)}
\end{eqnarray}

Substituting this in (\ref{cm}), we find the center of mass energy for Hayward
space-time:
\begin{eqnarray}
\left(\frac{E_{cm}}{\sqrt{2}m_{0}}\right)^{2} &=&  1 +\frac{E_{1}E_{2}}{{\cal H}(r)}-
\frac{Z_{1} Z_{2}}{{\cal H}(r)}
-\frac{L_{1}L_{2}}{r^2} ~.\label{cmHS}
\end{eqnarray}
Now we can see in the following diagram,  the variation of radial velocity
and CM energy for Hayward space-time.
\begin{figure}[t]
\begin{center}
{\includegraphics[width=0.45\textwidth]{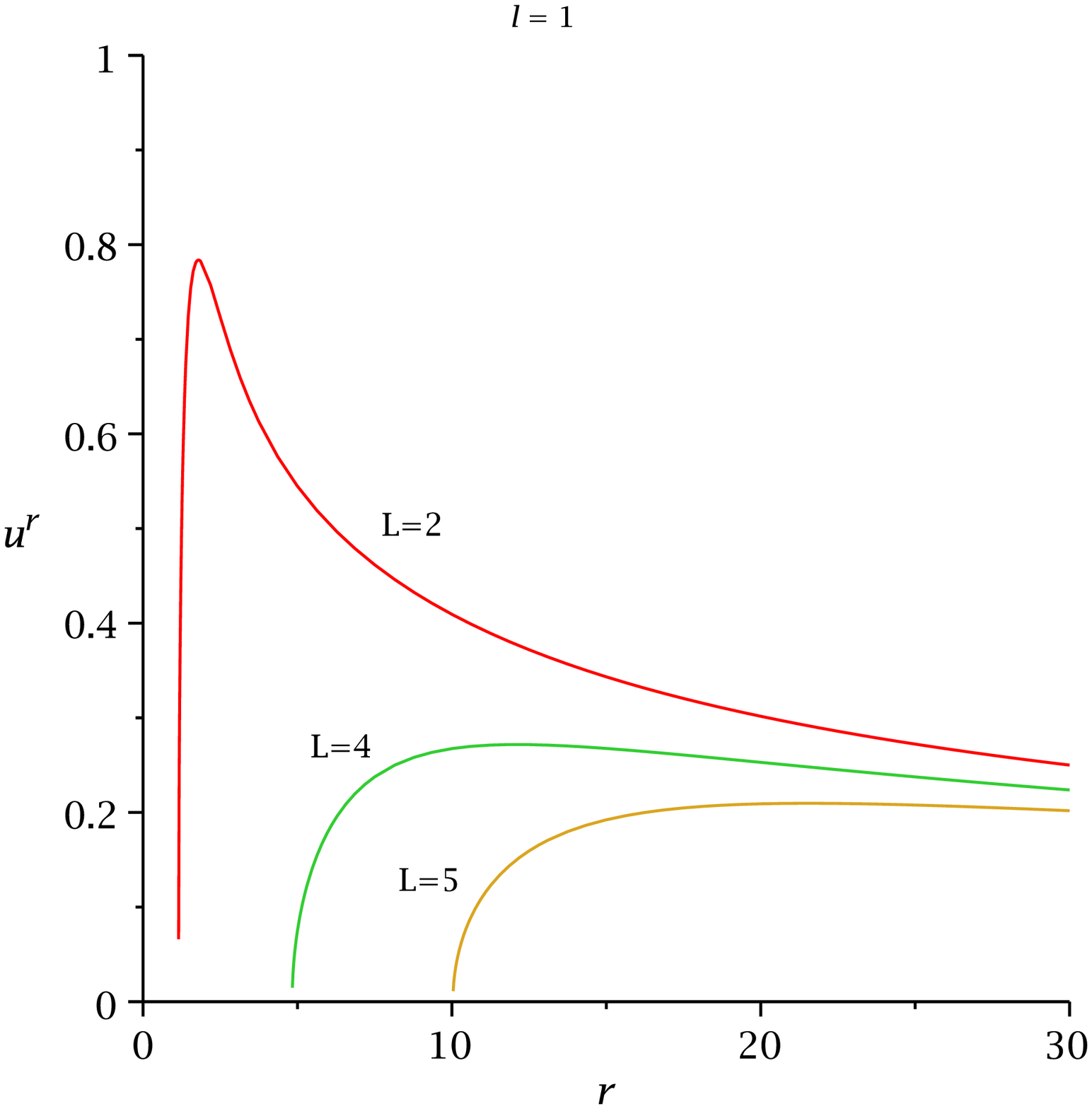}}
{\includegraphics[width=0.45\textwidth]{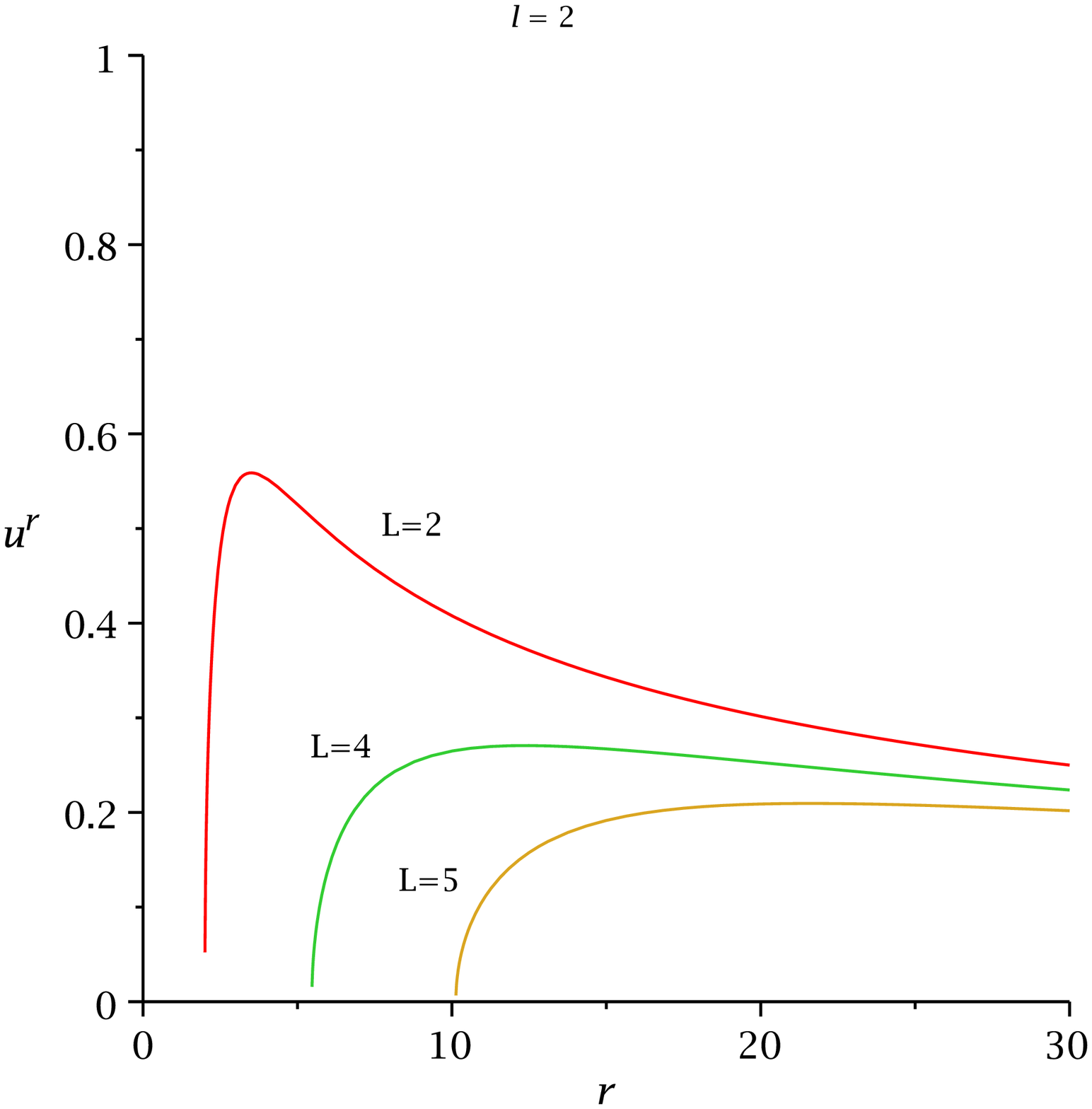}}
\end{center}
\caption{  The figure shows the variation of $\dot{r}$  with $r$ for Hayward
space-time. Here, $m=1$. \label{hayv}}
\end{figure}



\begin{figure}[t]
\begin{center}
{\includegraphics[width=0.45\textwidth]{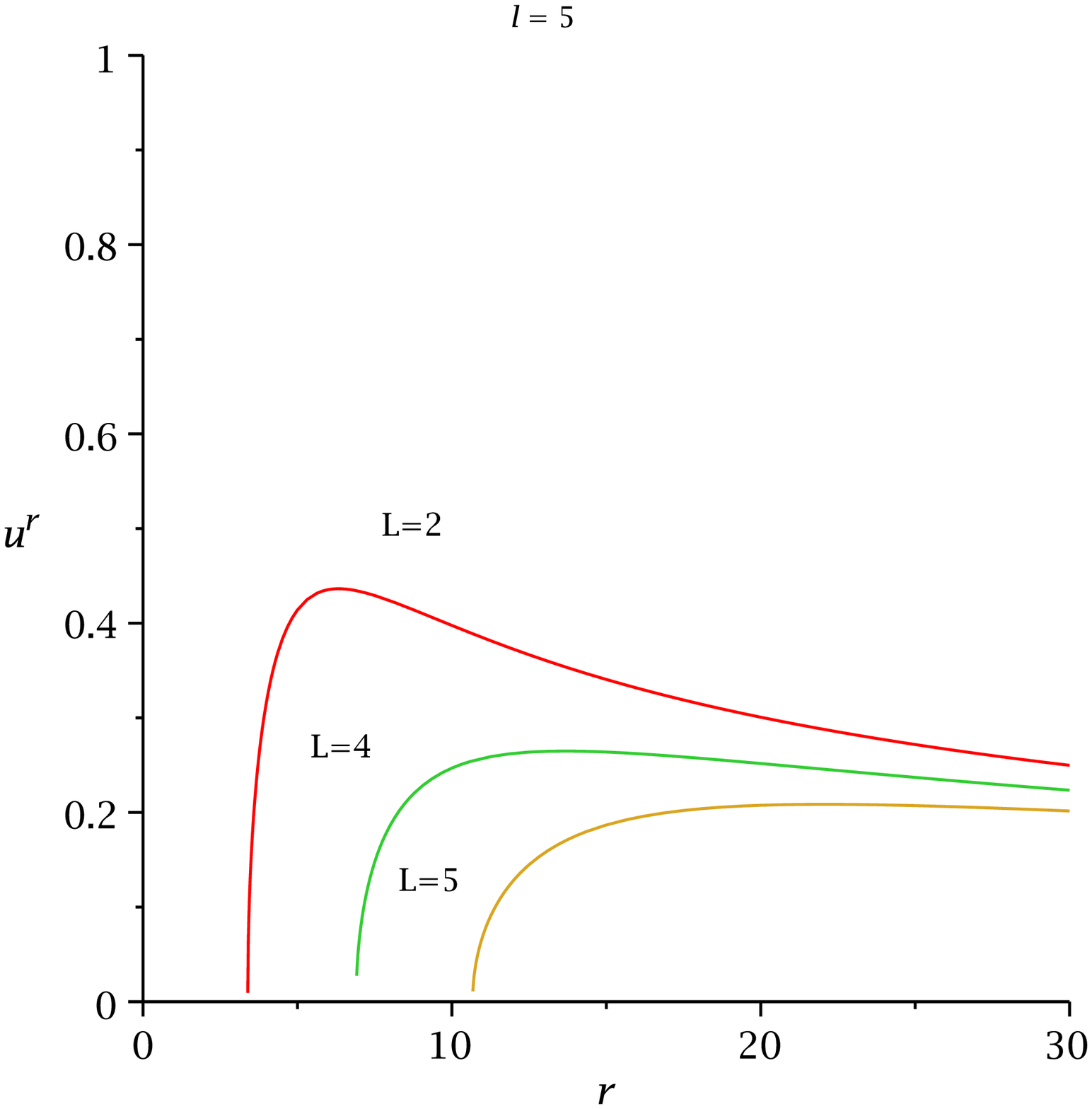}}
\end{center}
\caption{The figure shows the variation of $\dot{r}$  with $r$ for Hayward
space-time. Here, $m=1$. \label{hayv5}}
\end{figure}


\begin{figure}[t]
\begin{center}
{\includegraphics[width=0.45\textwidth]{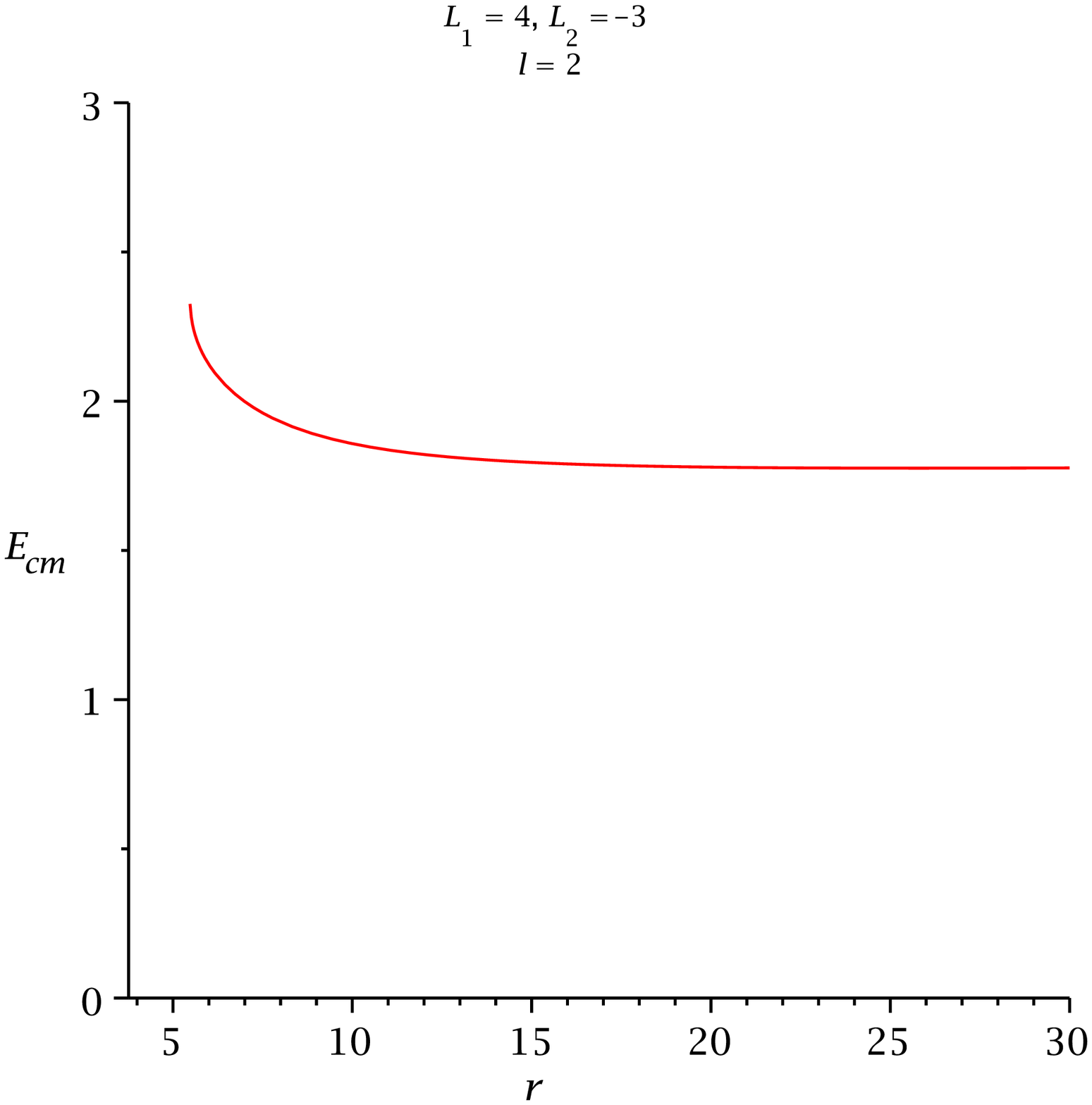}}
{\includegraphics[width=0.45\textwidth]{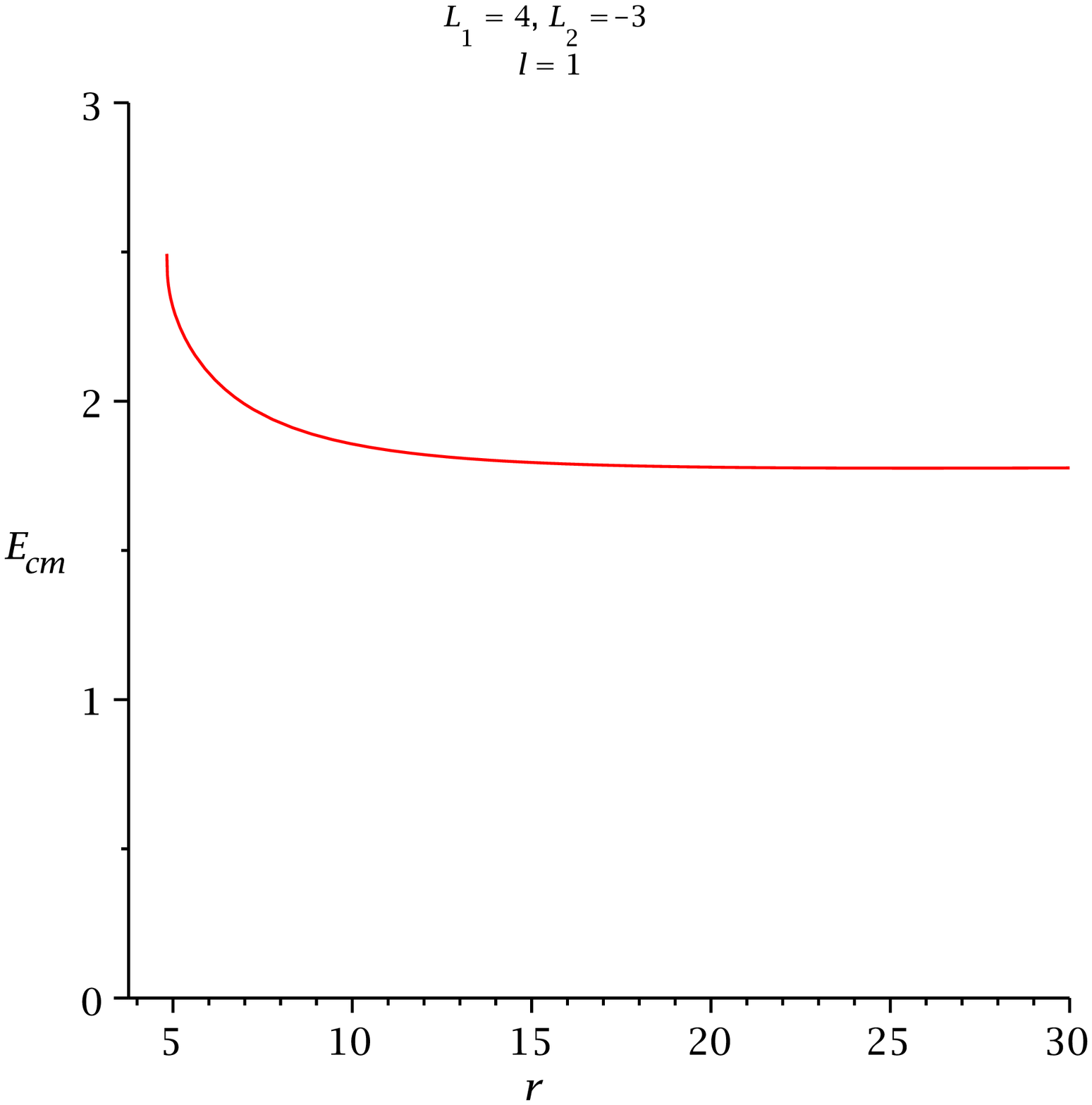}}
\end{center}
\caption{The figure shows the variation  of $E_{cm}$  with $r$ for Hayward black hole
with different values of angular momentum. \label{hayc1}}
\end{figure}



\begin{figure}[t]
\begin{center}
{\includegraphics[width=0.45\textwidth]{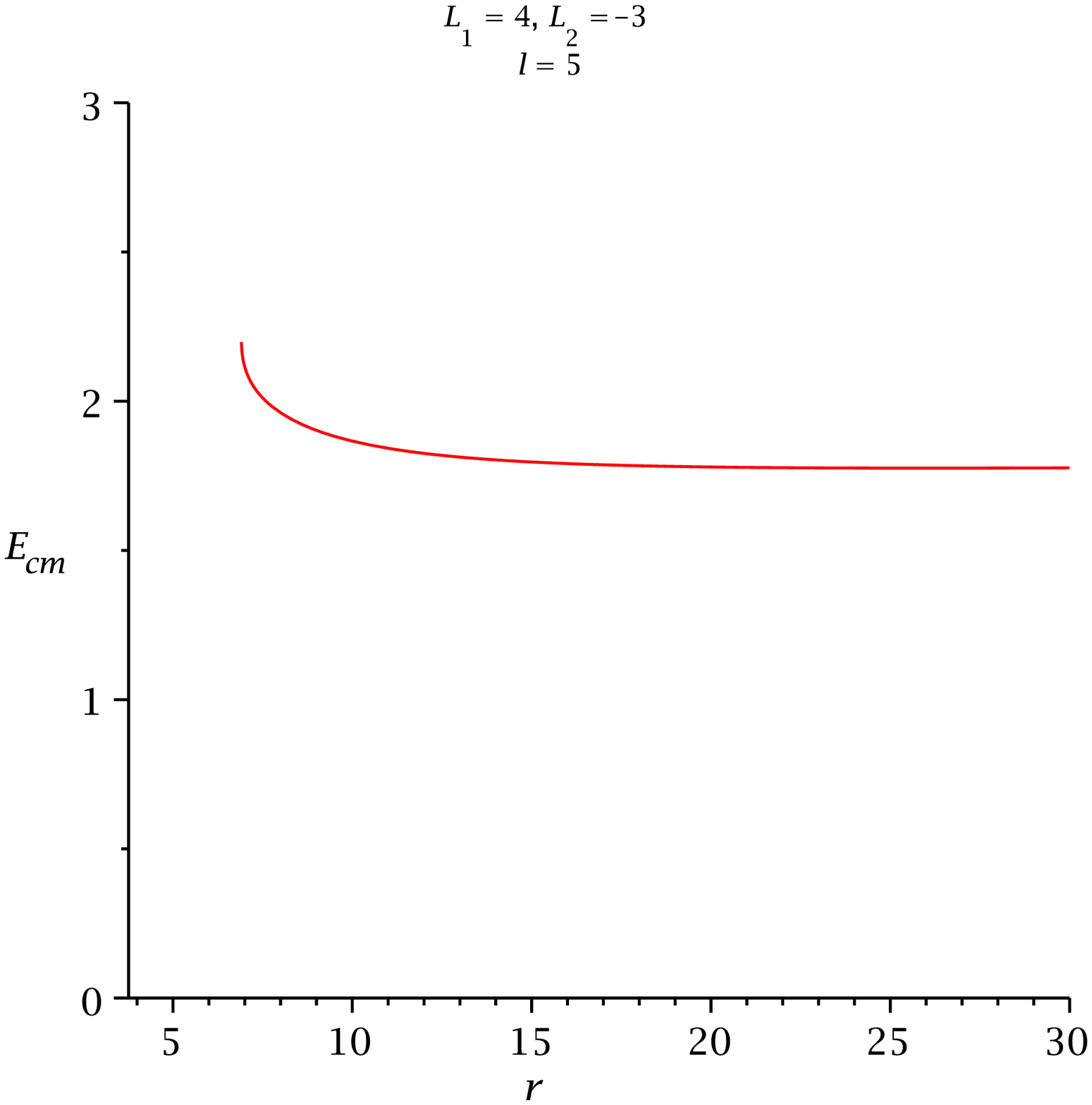}}
\end{center}
\caption{The figure shows the variation  of $E_{cm}$  with $r$ for Hayward black hole
with different values of angular momentum. \label{hayc3}}
\end{figure}


Taking, $E_{1}=E_{2}=1$ and substituting the value of ${\cal H}(r)$, we obtain
the CM energy near the event horizon ($r_{+}$) of the Hayward space-time:

\begin{eqnarray}
E_{cm}\mid_{r\rightarrow r_{+}} &=& \sqrt{2}m_{0}\sqrt{\frac{4r_{+}^2+(L_{1}-L_{2})^{2}}{2r_{+}^2}} ~.\label{cmahs}
\end{eqnarray}
where $r_{+}$ is  given in (\ref{hoha1}).

When we set $r_{+}=2m$,  we recover  the CM energy of the Schwarzschild black hole:
\begin{eqnarray}
E_{cm} &=& \sqrt{2}m_{0}\sqrt{\frac{16m^2+(L_{1}-L_{2})^{2}}{8m^2}} ~.\label{cmsch2}
\end{eqnarray}

The angular velocity of the Hayward space-time at the $r_{+}$ is given by
\begin{eqnarray}
\Omega_{H}=\frac{\dot{\phi}}{\dot{t}}=\sqrt{\frac{m(r_{0}^3-4ml^2)}{(r_{0}^3+2ml^2)^{\frac{5}{2}}}} ~.\label{omghs}
\end{eqnarray}
The critical angular momenta $L_{i}$ can be written as

\begin{eqnarray}
L_{i} &=& \frac{E_{i}}{\Omega_{H}} ~.\label{lih}
\end{eqnarray}
In the extremal cases, i.e.  $27l^2=16m^2$, the horizon is at $r_{0}=\frac{4}{3}m$ and
if one of the values of critical angular momenta diverge  we get the infinite amount of
CM energy, i.e.

\begin{eqnarray}
 E_{cm} & \longmapsto &  \infty   ~.\label{cmdivHs}
\end{eqnarray}

\section{Summary and Conclusions:}

In this work, we have demonstrated that the collision of two
neutral particles falling freely from rest at infinity in the
background  of the regular black holes.
Firstly, we have examined the particle acceleration and collision near the
infinite red-shift surface of the Bardeen space-time which is
the \emph{first} regular (singularity-free) black hole model in GR.
We proved  that the center of mass energy is arbitrarily large at
the infinite red-shift surface when the black hole is purely extremal.
For non-extremal Bardeen black hole, we found that the CM energy is
finite and depends upon the critical values of the angular momentum
parameter.

Secondly, we have examined the BSW mechanism near the
infinite red-shift surface of the ABG
black hole which is also  a regular black hole
space-time and singularity free solutions of the coupled system of a non-linear
electrodynamics and general relativity. We have showed that the center of mass energy
for this black hole also is arbitrarily large at the infinite red-shift surface
when it is in the purely extremal situation. For non-extremal ABG black hole, we
have seen that the CM energy is finite and depends upon the fine tuning condition
of the angular momentum parameter.

Finally, we have tested the BSW mechanism near the infinite red-shift surface
of the Hayward black hole which is also a singularity free solution in GR.
It is shown that these regular black holes
may act as natural particle accelerators with arbitrarily high CM energy
when the black hole is precisely extremal.

Moreover, in each cases, we have also studied the properties of equatorial circular
geodesic motion by extremization of  the effective potential for time-like circular orbits and null circular
orbits. We particularly emphasized on the  ISCO, MBCO and CPO of these regular black holes. These orbits
are useful to extract the information about the back ground geometry and they are also relevant to the astrophysical process.

Our conclusion is that for non-extremal regular space-time  the CM energy is
finite and depends upon  the angular momentum parameter. For extremal regular black holes, the CM energy
is unlimited due to the diverging values of the angular momentum parameter of the colliding particles.


\begin{thebibliography}{99}

\bibitem{bsw} M. Ba\~{n}ados , J. Silk , and S. M. West, \textit{Phys. Rev. Lett.} {\bf 103}, 111102 (2009).

\bibitem{berti} E. Berti  , V. Cardoso ,  L. Gualtieri , F. Pretorius  , and
 U. Sperhake, \textit{Phys. Rev. Lett.} {\bf 104}, 239001 (2009).

\bibitem{thorn} K. S. Thorn, \textit{Astrophys. J.} {\bf 191}, 507 (1974).

\bibitem{jacob} T. Jacobson  and  T. P. Sotiriou, \textit{Phys. Rev. Lett.}
{\bf 104}, 021101 (2010).

\bibitem{lake} ~ K. Lake, \ Phys. Rev. Lett. {\bf 104}, 211102;
{\bf 104}, 259903 (2010).

\bibitem{grib} A. ~ Grib , Y.~ Pavlov, \textit{Astropart. Phys.} {\bf 34}, 581 (2011).

\bibitem{grib1} A. ~ Grib , Y.~ Pavlov, \textit{Euro. Phys. Lett.}  {\bf 101}, 2004 (2013).

\bibitem{harada} T. ~ Harada , M.~ Kimura, \ Phys.~Rev.~D~{\bf 83}, 024002 (2011).

\bibitem{liu}  C.~ Liu , S.~ Chen , C.~ Ding , J.~ Jing, \textit{Phys. Letters B }{\bf 701}, 285-290 (2011).

\bibitem{li} Y.~ Li , J.~ Yang , Y.~ Li , S. Wei , Y. Liu ~, \textit{Class. Quantum Gravity} {\bf 28}, 225006 (2011).

\bibitem{zong} C.~ Zhong , S. ~ Gao, \textit{JETP Letters} {\bf 94}, 589 (2011).

\bibitem{said} J. L.~ Said , K. Z. Adami, \ Phys.~Rev.~D~{\bf 83}, 104047 (2011).

\bibitem{piran} M. Bejger,  T. Piran,  M. Abramowicz, F. H{\aa}kanson, \textit{Phys. Rev.
    Lett.} {\bf 109} 121101 (2012).

\bibitem{wei1} S. W. Wei, Y. X. Liu ~, H. Guo, C. E. Fu ~, Phys.~Rev.~D~{\bf 82}, 103005
(2010).

\bibitem{wei} S. Wei ~, Y. Liu ~, H. Li ~, F. Chen, \ JHEP {\bf 1012}, 066 (2010).

\bibitem{zaslav} O. Zaslavskii ~, \ JETP Lett. {\bf 92}, 571 (2010).

\bibitem{zhu} Y. Zhu ~,  S. Wu ~, Y. Jiang ~,  G. Yang ~, \ Phys.~Rev.~D~{\bf 84}, 123002, 043006 (2011).

\bibitem{huss} I. Hussain,  \ Modern Physics Letters A {\bf 27}, 1250068 (2012).

\bibitem{frolov} V. P. Frolov, \ Phys. Rev., {\bf D 86}, 044040 (2012).


\bibitem{sharif} M. Sharif , N. Haider, \ Astrophys Space Sci., {DOI: 10.1007/s10509-013-1424-3} (2013).

\bibitem{mc} S. McWilliams~, \ Phys. Rev. Lett., {\bf 110}, 011102 (2012).

\bibitem{gala} A. Galajinsky, \ Phys. Rev. , {\bf D 88}, 027505 (2013).

\bibitem{tur} A. Tursunov , M. Kolo\v{s} , A. Abdujabbarov , B. Ahmedov  and
Z. Stuchl\'{i}k , \ Phys. Rev., {\bf D 88}, 124001 (2013).

\bibitem{fernando} S. Fernando, \ Gen. Rel. Gravit.  {\bf 46} {1634} (2014).






\bibitem{patil5}  M.~ Patil , P. ~Joshi~,  \ Phys. Rev. D {\bf 86}, 044040 (2012).


\bibitem{patil7} A. N. Chowdhury ~, M. Patil ~, D. Malafarina , P. S. Joshi~, \ Phys. Rev.  {\bf  D 85}, 104031 (2012).

\bibitem{pp3} P. Pradhan, \ Astrophys Space Sci., {DOI: 10.1007/s 10509-014-1896-9} (2014).

\bibitem{bard} J. Bardeen, \ Conference Proceedings in  GR5,  Tiflis, U.S.S.R., (1968).

\bibitem{abg1}  E. Ay\'{o}n-Beato , A. Garc\'{i}a , \ Phys. Rev. Lett. {\bf 80}, 5056 (1998).

\bibitem{hay}  S. A. Hayward, { Phys. Rev. Lett.} {\bf 96}, 031103 (2006).

\bibitem{abg}  E. Ay\'{o}n-Beato ,  A. Garc\'{i}a, \ Physics Letters {\bf B 493}
, 149-152 (2000).

\bibitem{borde} A. Borde, \ Phys. Rev., {\bf  D 55}, 7615 (1997).

\bibitem{ansol} S. Ansoldi, {Arxiv: 0802.0330} [gr-qc] (2008).

\bibitem{lb} L. Balart, { Physics Letters } {\bf B 687} 280-285 (2010).

\bibitem{bose} S. Bose, N. Dadhich, \ Phys. Rev., {\bf D 60}, 064010 (1999).

\bibitem{nora} N. Bret\'{o}n , \ Gen. Rel. Gravit.  {\bf 37(4)} {643-650} (2005).

\bibitem{bala}  L. Balart ,  E. C. Vagenas , { Physics Letters } {\bf B 730} 14-17 (2014).

\bibitem{eva} A. Garc\'{i}a ,  E. Hackman ,  J. Kunz ,  C. L\"{a}mmerzahl  and
A. Macias, `` Motion of test particles in a regular black hole space-time'', arXiv:1306.2549 (2013).

\bibitem{eiro} E. F. Eiroa , C. M. Sendra, {Classical and Quantum Gravity}, {\bf 28}(8) 085008 (2011).

\bibitem{zhou} S. Zhou , J.  Chen, and  Y. Wang , {International Journal of  Modern Physics } {\bf D 21}, 1250077 (2011).

\bibitem{sch} S. Chandrashekar , {\it The Mathematical Theory of Black Holes}, Clarendon Press, Oxford (1983).

\bibitem{hart} J. B. Hartle , {\it Gravity-An Introduction To Einstein's General Relativity }, Benjamin Cummings (2003).

\bibitem{bau} A. N. Baushev, {\ Int. J.   Mod. Phys. D} {\bf 18}, 1195 (2009).

\bibitem{abg2} E. Ay\'{o}n-Beato, A. Garc\'{i}a, \ Physics Letters {\bf B 464}, 25-29
(1999).

\bibitem{abg3} E. Ay\'{o}n-Beato  , A. Garc\'{i}a, \ Gen. Rel. Gravit. {\bf 629}, 31
(1999).

\bibitem{abg4}  E. Ay\'{o}n-Beato ,  A. Garc\'{i}a, \ Gen. Rel. Gravit. {\bf 635}, 37
(2005).

\bibitem{bron} K. A. Bronnikov , J. C. Fabris, {\ Phys. Rev. Lett.} {\bf 96}, 251101 (2006).

\end{thebibliography}
\end{document}